\documentclass[12pt]{article}
\usepackage{amsmath,amssymb,amsfonts}

\def\op#1{\mathop{{\it\fam0} #1}\limits}

\newcommand{\ccG}{{\mathfrak g}}

\def\op#1{\mathop{\fam0 #1}\limits}

\newcommand{\id}{{\mathrm{Id}\,}}

\newcommand{\cA}{{\mathcal A}}
\newcommand{\cT}{{\mathcal T}}

\newcommand{\cD}{{\mathcal D}}
\newcommand{\cR}{{\mathcal R}}
\newcommand{\cV}{{\mathcal V}}
\newcommand{\cC}{{\mathcal{C}\ell}}
\newcommand{\lC}{{\mathcal C}}
\newcommand{\cL}{{\mathcal L}}
\newcommand{\cE}{{\mathcal E}}

\newcommand{\cG}{{\mathcal G}}

\newcommand{\cS}{{\mathcal S}}

\newcommand{\cJ}{{\mathcal J}}
\newcommand{\cK}{{\mathcal K}}
\newcommand{\cZ}{{\mathcal Z}}

\newcommand{\bL}{{\mathbf L}}

\newcommand{\bF}{{\mathbf F}}

\newcommand{\rL}{{\mathrm L}}

\newcommand{\al}{\alpha}

\newcommand{\bt}{\beta}
\newcommand{\dl}{\delta}
\newcommand{\la}{\lambda}

\newcommand{\om}{\omega}

\newcommand{\m}{\mu}
\newcommand{\n}{\nu}

\newcommand{\g}{\gamma}
\newcommand{\G}{\Gamma}
\newcommand{\thh}{\theta}

\newcommand{\vt}{\vartheta}

\newcommand{\si}{\sigma}
\newcommand{\Si}{\Sigma}
\newcommand{\w}{\wedge}
\newcommand{\wt}{\widetilde}
\newcommand{\wh}{\widehat}
\newcommand{\ol}{\overline}
\newcommand{\dr}{\partial}
\newcommand{\ar}{\op\longrightarrow}
\newcommand{\ot}{\otimes}
\newcommand{\ap}{\approx}
\newcommand{\ve}{\varepsilon}

\newcommand{\beq}{\begin{equation}}
\newcommand{\eeq}{\end{equation}}
\newcommand{\ben}{\begin{eqnarray}}
\newcommand{\een}{\end{eqnarray}}
\newcommand{\be}{\begin{eqnarray*}}
\newcommand{\ee}{\end{eqnarray*}}

\newcounter{eqalph}
\newcounter{equationa}

\let\ssection=\section
\renewcommand{\section}{\setcounter{equation}{0}\ssection}

\newcounter{example}[section]
\newcounter{remark}[section]
\newcounter{theorem}[section]
\newcounter{proposition}[section]
\newcounter{lemma}[section]
\newcounter{corollary}[section]
\newcounter{definition}[section]

\def\theremark{\arabic{section}.\arabic{remark}}
\def\thetheorem{\arabic{section}.\arabic{theorem}}

\def\thedefinition{\arabic{section}.\arabic{definition}}

\newenvironment{example}{\refstepcounter{remark}\medskip\noindent{\bf
Example \theremark:} }{$\Box$ \medskip}
\newenvironment{remark}{\refstepcounter{remark}\medskip\noindent{\bf
Remark \theremark:} }{$\Box$\medskip}
\newenvironment{theorem}{\refstepcounter{theorem}
\medskip\noindent{\sc Theorem \thetheorem}:}{$\Box$\medskip}

\newenvironment{lemma}{\refstepcounter{theorem}\medskip\noindent{\sc
Lemma \thetheorem}:}{ $\Box$\medskip }
\newenvironment{corollary}{\refstepcounter{theorem}\medskip\noindent{\sc
Corollary \thetheorem}:}{$\Box$\medskip}

\newcommand{\mar}[1]{}

\begin{document}
\hbox{}

\begin{center}

{\large \textbf{Lecture on Gauge Gravitation Theory. \\ Gravity as
a Higgs Field}}

\bigskip

\textbf{G. Sardanashvily}

\medskip

Moscow State University, Russia

Lepage Research Institute, Czech Republic

\bigskip

20th International Summer School on Global Analysis and its
Applications \textbf{"General Relativity: 100 years after
Hilbert"} (Star\'a Lesn\'a, Slovakia, 2015)

\end{center}

\noindent \textbf{Abstract.} Gravitation theory is formulated as
gauge theory on natural bundles with spontaneous symmetry breaking
where gauge symmetries are general covariant transformations,
gauge fields are general linear connections, and Higgs fields are
pseudo-Riemannian metrics.

\tableofcontents

\section{Introduction}

Theory of classical fields admits a comprehensive mathematical
formulation in the geometric terms of smooth fibre bundles over
$X$ \cite{book09,sard08,book13}. For instance, Yang--Mills gauge
theory is theory of principal connections on principal bundles.

Gravitation theory on a world manifold $X$ is formulated as gauge
theory on natural bundles over $X$ which admit general covariant
transformations as the canonical functorial lift of
diffeomorphisms of their base $X$ \cite{book09,sard11}. This is
metric-affine gravitation theory where gauge fields are general
linear connections (Section 5), and  a metric gravitational field
is treated as a classical Higgs field responsible for reducing a
structure group of natural bundles to a Lorentz group (Section 6).
The underlying physical reason of this reduction is both the
geometric Equivalence principle and the existence of Dirac spinor
fields. Herewith, a structure Lorentz group always is reducible to
its maximal compact subgroup of spatial rotations that provides a
world manifold $X$ with an associated space-time structure and
metric space topology (Section 7).

Spontaneous symmetry breaking is a quantum phenomenon when
automorphism of a quantum algebra need not preserve its vacuum
state \cite{pref,sard15}. In this case, we have inequivalent
vacuum states of a quantum system which are classical objects. The
physical nature of gravity as a Higgs field is characterized by
the fact that, given different gravitational fields, the
representations (\ref{L4'}) of holonomic coframes $\{dx^\mu\}$ on
a world manifold $X$ by $\gamma$-matrices acting on spinor fields
are non-equivalent and, consequently, the Dirac operators in the
presence of different gravitational field fails to be equivalent,
too (Section 10). This fact motivates us to think that a metric
gravitational field is not quantized in principle.

\section{History}

A first model of gauge gravitation theory was suggested by
R.Utiyama \cite{uti} in 1956 just two years after the birth of
gauge theory itself. He was first who generalized the original
gauge model of Yang and Mills for $SU(2)$ to an arbitrary symmetry
Lie group and, in particular, to a Lorentz group in order to
describe gravity. However, he met the problem of treating general
covariant transformations and a pseudo-Riemannian metric which had
no partner in Yang--Mills gauge theory.

To eliminate this drawback, representing a tetrad gravitational
field as a gauge field of a translation subgroup of a Poincar\'e
group was attempted because, by analogy with gauge potentials in
Yang--Mills gauge theory, the indices $a$ of a tetrad field
$h^a_\mu$ were treated as those of a translation group (see
\cite{blag,volume,cap,hehl,iva,obukh,sard06} and references
therein). Since the Poincar\'e group comes from the
Wigner--In\"onii contraction of de Sitter groups $SO(2,3)$ and
$SO(1,4)$ and it is a subgroup of a conformal group, gauge
theories on fibre bundles $Y\to X$ with these structure groups
were also considered \cite{gotz,hay,ivan,lecl,stell,tseytl}.
Because these fibre bundles fail to be natural, the lift of the
group $Diff(X)$ of diffeomorphisms of $X$ onto $Y$ should be
defined \cite{lord,lord2}. In a general setting, one can study a
gauge theory on a fibre bundle with the typical fibre $\mathbb
R^n$ and the topological structure group $Diff(\mathbb R^n)$ or
its subgroup of analytical diffeomorphisms \cite{bor,kirsch}. The
Poincar\'e gauge theory also is generalized to the higher $s$-spin
gauge theory of tensor coframes and generalized Lorentz
connections, which satisfy certain symmetry, skew symmetry and
traceless conditions \cite{vasil1}.

A problem however is that that a non-linear (translation) summand
of an affine connection (Section 12) is a soldering form, but
neither frame (vierbein) field nor tetrad field. The latter thus
has no the status of a gauge field \cite{iva,sard83,sard11}. At
the same time, a translation part of an affine connection on
$\mathbb R^3$ characterizes an elastic distortion in gauge theory
of dislocations in continuous media \cite{kad,maly}. A similar
gauge model of hypothetic deformations of a world manifold has
been developed and, in particular, they may be responsible for the
so called "fifth force" \cite{sard87,sard90,sardz}.

At the same time, gauge theory in a case of spontaneous symmetry
breaking also contains classical Higgs fields, besides the gauge
and matter ones
\cite{book09,iva,keyl,nik,sard92,higgs,pref,sard14,tra}.
Therefore, basing on the mathematical definition of a
pseudo-Riemnnian metric, we have formulated gravitation theory as
gauge theory with a Lorentz reduced structure where a metric
gravitational field is treated as a Higgs field
\cite{book09,iva,sard80,sardz,sard06,sard11}.

\section{Main Theses}

Gauge gravitation theory in comparison with the Yang--Mills one
possesses the following peculiarities \cite{book09,sard11}.

$\bullet$ Gauge symmetries of gravitation theory are general
covariant transformations which are not vertical automorphisms of
principal bundles in Yang--Mills gauge theory.

$\bullet$ Gauge gravitation theory necessarily is theory with
spontaneous symmetry breaking in the presence of the corresponding
Higgs fields. Since gauge symmetries of gravitation theory are
general covariant transformations, but not vertical automorphisms
of fibre bundles, these Higgs fields, unlike Higgs fields in
Yang--Mills gauge theory, are dynamic variables.

$\bullet$ In comparison with Yang--Mills gauge theory, e.g., the
Standard Model of particle physics \cite{nov,SM}, matter fields in
gauge gravitation theory admits only exact symmetries. These are
Dirac spinor fields with Lorentz spin symmetries, and there is a
problem of describing their general covariant transformations.

$\bullet$ The gauge invariance gauge gravitation theory under
general covariant transformation leads to a conservation law of an
energy-momentum symmetry current, but not the Noether one in
Yang--Mills gauge theory.

Studying gauge gravitation theory, we believe reasonable to
require that it incorporates Einstein's General Relativity and,
therefore, it should be based on Relativity and Equivalence
Principles reformulated in the fibre bundle terms
\cite{iva81,iva}.

In these terms, Relativity Principle states that gauge symmetries
of classical gravitation theory are general covariant
transformations \cite{book09,sard11}.

Let $\pi:Y\to X$ be a smooth fibre bundle. Any automorphism
$(\Phi,f)$ of $Y$, by definition, is projected as $\pi \circ \Phi=
f\circ \pi$ onto a diffeomorphism $f$ of its base $X$. The
converse is not true.

A fibre bundle $Y\to X$ is called the natural bundle if there
exists a monomorphism
\be
\mathrm{Diff} X\ni f\to\wt f\in \mathrm{Aut} Y
\ee
of the group of diffeomorphisms of $X$ to the group of bundle
automorphisms of $Y\to X$. Automorphisms $\wt f$ are called
general covariant transformations of $Y$.

Accordingly, there is the functorial lift of any vector field
$\tau$ on $X$ to a vector field $\ol\tau$ on $Y$ such that
$\tau\mapsto\ol\tau$ is a monomorphism of the Lie algebra $\cT(X)$
of vector field on $X$ to that $\cT(T)$ of vector fields on $Y$.
This functorial lift $\ol\tau$ is an infinitesimal generator of a
local one-parameter group of local general covariant
transformations of $Y$.

As was mentioned above, general covariant transformations differ
from gauge symmetries of Yang--Mills gauge theory which are
vertical automorphisms of principal bundles. Fibre bundles
possessing general covariant transformations constitute the
category of so called natural bundles \cite{kol,terng}.

The tangent bundle $TX$ of $X$ exemplifies a natural bundle. Any
diffeomorphism $f$ of $X$ gives rise to the tangent automorphisms
$\wt f=Tf$ of $TX$ which is a general covariant transformation of
$TX$. The associated principal bundle is a fibre bundle $LX$ of
linear frames in the tangent spaces to $X$. It also is a natural
bundle. Moreover, all fibre bundles associated to $LX$ are natural
bundles, but not they are only. Principal connections on $LX$
yield linear connections on the tangent bundle $TX$ and other
associated bundles over a world manifold. They are called the
world connections.

Following Relativity Principle, one thus should develop
gravitation theory as a gauge theory of principal connections on a
principal frame bundle $LX$ over an oriented four-dimensional
connected smooth manifold $X$, called the world manifold.

\begin{remark}
Smooth manifolds throughout are assumed to be Hausdorff
second-countable (consequently, locally compact and paracompact)
topological spaces.
\end{remark}

Equivalence Principle reformulated in geometric terms requires
that the structure group
\mar{gg1}\beq
GL_4=GL^+(4,\mathbb R) \label{gg1}
\eeq
of a frame bundle $LX$ and associated bundles over a world
manifold $X$ is reducible to a Lorentz group $SO(1,3)$
\cite{iva,sardz,sard11}. It means that these fibre bundles admit
atlases with $SO(1,3)$-valued transition functions or,
equivalently, that there exist principal subbubdles of $LX$ with a
Lorentz structure group. This is the case of spontaneous symmetry
breaking in classical gauge theory.

As was mentioned above, spontaneous symmetry breaking is a quantum
phenomenon when automorphism of a quantum algebra need not
preserve its vacuum state \cite{pref,sard15}. In this case, we
have inequivalent vacuum states of a quantum system which are
classical objects. For instance, spontaneous symmetry breaking in
Standard Model of particle physics is ensured by the existence of
a constant vacuum Higgs field which takes a value into the
quotient $G/H$ of a broken symmetry group $G$ by the exact one $H$
\cite{nov,SM}.

Therefore, classical gauge theory on principal bundles with
spontaneous symmetry breaking also is considered. This phenomenon
is characterized as a reduction of a structure Lie group $G$ of a
principal bundle $P\to X$ to its closed Lie subgroup $H$
\cite{book09,higgs,pref,sard14}. One refers to the following
reduction theorem \cite{kob}.

\begin{theorem} \label{red} \mar{red}
There exists one-to-one correspondence between the principal
$H$-subbundles $P^h$ of $P$ and the global sections $h$ of the
quotient bundle $P/H\to X$ with a typical fibre $G/H$.
\end{theorem}

These global sections are treated as classical Higgs fields
\cite{book09,higgs,sard14}.

Accordingly, in gauge gravitation theory based on Equivalence
Principle, there is one-to-one correspondence between the Lorentz
principal subbundles of a frame bundle $LX$ (called the Lorentz
reduced structures) and the global sections of the quotient bundle
\mar{b3203}\beq
\Si_{\mathrm{PR}}= LX/SO(1,3),\label{b3203}
\eeq
which are pseudo-Riemannian metrics on a world manifold. In
Einstein's General Relativity, they are identified with
gravitational fields.

Thus, gauge gravitation theory leads us to metric-affine
gravitation theory whose dynamic variables are linear world
connections and pseudo-Riemannian metrics on a world manifold $X$
(Section 8). They are treated as gauge fields and Higgs fields,
respectively \cite{book09,sard11}.

There is the extensive literature on metric-affine gravitation
theory \cite{blag,hehl,heh07,obukh}. However, one often formulates
it as gauge theory of affine connections, that is incorrect
(Section 12). Let us also emphasize that gauge gravitation theory
deals with general linear connections which need not be the
Lorentz connections.

The character of gravity as a Higgs field responsible for
spontaneous breaking of general covariant transformations is
displayed as follows. Given different gravitational fields, the
representations (\ref{L4'}) of holonomic coframes $\{dx^\m\}$ by
$\g$-matrices acting on spinor fields are inequivalent (Remark
\ref{gg20}). In particular, it follows that a Dirac spinor field
can be considered only in a pair with a certain gravitational
field. A total system of such pairs is described by sections of
the composite bundle $S\to \Si_{\mathrm T}\to X$ (\ref{ggz}),
where $S\to \Si_{\mathrm T}$ is a spinor bundle.

Being reduced to a Lorentz group, a structure group of a frame
bundle $LX$ also is reduced to a maximal compact subgroup $SO(3)$
of $SO(1,3)$. The associated Higgs field is a spatial distribution
which defines a space-time structure on a world manifold $X$
(Section 7).

Since general covariant transformations are symmetries of a
metric-affine gravitation Lagrangian, the corresponding
conservation law holds (Section 9). It is an energy-momentum
conservation law. Because general covariant transformations are
gauge transformations depending on derivatives of gauge
parameters, the corresponding energy-momentum current reduces to a
superpotential \cite{book09,sard09,sard11}. This is the
generalized Komar superpotential (\ref{K3}) .

\section{Natural bundles}

Let $\pi:Y\to X$ be a smooth fibre bundle coordinated by $(x^\la,
y^i)$. Given a one-parameter group $(\Phi_t,f_t)$ of automorphisms
of $Y$, its infinitesimal generator is a projectable vector field
\be
u=\tau^\la(x^\m)\dr_\la + u^i(x^\m,y^j)\dr_i
\ee
on $Y$ which is projected onto a vector field
$\tau=\tau^\la\dr_\la$ on $X$, whose flow is a one-parameter group
$(f_t)$ of diffeomorphisms of $X$. Conversely, let
$\tau=\tau^\la\dr_\la$ be a vector field on $X$. Its lift to some
projectable vector field on $Y$ always exists. For
instance, given a connection
\be
\G=dx^\la\ot(\dr_\la +\G^i_\la(x^\m,y^j)\dr_i)
\ee
on $Y\to X$, a vector field $\tau$ on $X$ gives rise to a
horizontal vector field
\be
\G\tau=\tau\rfloor\G=\tau_\la (\dr_\la+\G^i_\la\dr_i)
\ee
on $Y$. The horizontal lift $\tau\to\G\tau$ yields a monomorphism
of a $C^\infty(X)$-module $\cT(X)$ of vector fields on $X$ to a
$C^\infty(Y)$-module $\cT(Y)$ of vector fields on $Y$, but this
monomorphism is not a Lie algebra morphism, unless $\G$ is flat.

We address the category of \textbf{natural bundles} $Y\to X$
admitting the functorial lift $\wt\tau$ onto $Y$ of any vector
field $\tau$ on $X$ such that $\tau\to\ol\tau$ is a Lie algebra
monomorphism $\cT(X)\to \cT(T)$,
$[\wt\tau,\wt\tau']=\wt{[\tau,\tau']}$ \cite{book09,kol,terng}.
This functorial lift $\wt\tau$, by definition, is an infinitesimal
generator of a local one-parameter group of \textbf{general
covariant transformations} of $Y$.

Natural bundles are exemplified by tensor products
\mar{sp20}\beq
T=(\op\ot^m TX)\ot(\op\ot^k T^*X) \label{sp20}
\eeq
of the tangent $TX$ and cotangent $T^*X$ bundles of $X$. Given a coordinate atlas $(x^\m)$ of $X$, the tangent
bundle $\pi_X:TX\to X$ is provided with holonomic bundle
coordinates
\be
(x^\m,\dot x^\m), \qquad \dot x'^\m =\frac{\dr x'^\m}{\dr
x^\nu}\dot x^\nu,
\ee
where $(\dot x^\m)$ are fibre coordinates with respect to
holonomic frames $\{\dr_\m\}$. Accordingly, the tensor bundle
(\ref{sp20}) is endowed with holonomic bundle coordinates
$(x^\m,x^{\al_1\cdots\al_m}_{\bt_1\cdots\bt_k})$, where
\be
(x^\m,\dot x_\m), \qquad \dot x'_\m =\frac{\dr x^\nu}{\dr
x_\m}\dot x_\nu,
\ee
are those on the cotangent bundle $T^*X$ of
$X$. Then given a vector field $\tau$ on $X$, its functorial lift
onto the tensor bundle (\ref{sp20}) takes a form
\be \wt\tau =
\tau^\m\dr_\m + [\dr_\nu\tau^{\al_1}\dot
x^{\nu\al_2\cdots\al_m}_{\bt_1\cdots\bt_k} + \ldots
-\dr_{\bt_1}\tau^\nu \dot
x^{\al_1\cdots\al_m}_{\nu\bt_2\cdots\bt_k} -\ldots]\dot \dr
_{\al_1\cdots\al_m}^{\bt_1\cdots\bt_k}, \qquad \dot\dr_\la =
\frac{\dr}{\dr\dot x^\la}.
\ee

Tensor bundles over a world manifold $X$ have the structure group
$GL_4$ (\ref{gg1}). An associated principal bundle is the above
mentioned frame bundle $LX$. Its (local) sections are called
\textbf{frame (vierbein) fields}. Given a holonomic atlas of the
tangent bundle $TX$, every element $\{H_a\}$ of a frame bundle
$LX$ takes a form $H_a=H^\m_a\dr_\m$, where $H^\m_a$ is a matrix
of the natural representation of a group $GL_4$ in $\mathbb R^4$.
These matrices constitute bundle coordinates
\be
(x^\la, H^\m_a), \qquad H'^\m_a=\frac{\dr x'^\m}{\dr
x^\la}H^\la_a,
\ee
on $LX$ associated to its holonomic atlas
\mar{tty}\beq
\Psi_T=\{(U_\iota, z_\iota=\{\dr_\m\})\}, \label{tty}
\eeq
given by local frame fields $z_\iota=\{\dr_\m\}$.

A frame bundle $LX$ is equipped with a canonical $\mathbb
R^4$-valued one-form
\mar{b3133'}\beq
\thh_{LX} = H^a_\m dx^\m\ot t_a,\label{b3133'}
\eeq
where $\{t_a\}$ is a fixed basis for $\mathbb R^4$ and $H^a_\m$ is
the inverse matrix of $H^\m_a$.

A frame bundle $LX\to X$ is natural. Indeed, any diffeomorphism
$f$ of $X$ gives rise to an automorphism
\mar{025}\beq
\wt f: (x^\la, H^\la_a)\to (f^\la(x),\dr_\m f^\la H^\m_a)
\label{025}
\eeq
of $LX$ which is its general covariant transformation. Given a
(local) one-parameter group of diffeomorphisms of $X$ and its
infinitesimal generator $\tau$, the lift (\ref{025}) yields a
functorial lift
\be
\wt\tau=\tau^\m\dr_\m +\dr_\nu\tau^\al H^\nu_a\frac{\dr}{\dr
H^\al_a}
\ee
onto $LX$ of a vector field $\tau$ on $X$ which is defined by the
condition $\bL_{\wt\tau}\thh_{LX}=0$.

Let $Y=(LX\times V)/GL_4$ be an $LX$-associated bundle with a
typical fibre $V$. It admits a lift of any diffeomorphism $f$ of
its base to an automorphism
\be
f_Y(Y)=(\wt f(LX)\times V)/GL_4
\ee
of $Y$ associated to the principal automorphism $\wt f$
(\ref{025}) of a frame bundle $LX$. Thus, all bundles associated
to a frame bundle $LX$ are natural bundles.

\begin{remark} \label{gg3} \mar{gg3}
In a general setting, one also considers the total group Aut$(LX)$
of automorphisms of a frame bundle $LX$ \cite{hehl}. Such an
automorphism is the composition of some general covariant
transformation and a vertical automorphism of $LX$, which is a
non-holonomic frame transformation. Subject to vertical
automorphisms, the tangent bundle $TX$ is provided with
non-holonomic frames $\{\vt_a\}$ and the corresponding bundle
coordinates $(x^\m,y^a)$. A problem is that Lagrangians of
gravitation theory which factorize through the Ricci tensor
(\ref{ric}), e.g. the Hilbert--Einstein Lagrangian (\ref{10221})
are not invariant under non-holonomic frame transformations (see
Remark \ref{gg5} and Example \ref{gg10}). To overcome this
difficulty, one can additionally introduce frame
$\vt_a=\vt_a^\m\dr_\m$ (or coframe $\vt^a=\vt^a_\m dx^\m$) fields,
which are sections of a frame bundle $LX$. These sections are
necessarily local, unless $LX$ is a trivial bundle, i.e., $X$ is a
parallelizable manifold (Remark \ref{gg2}). In particular, this is
the case of theory of teleparallel gravity \cite{cai,obu}.
\end{remark}

\section{World connections}

Let $TX$ be the tangent bundle of a world manifold $X$. With
respect to holonomic coordinates $(x^\la,\dot x^\la)$, a linear
connection on $TX$ takes a form
\mar{B}\beq
\G= dx^\la\otimes (\dr_\la +\G_\la{}^\m{}_\n \dot x^\n
\dot\dr_\m). \label{B}
\eeq
It is called a \textbf{linear world connection} on $X$. Since $TX$
is associated to a frame bundle $LX$, every linear connection
(\ref{B}) is associated to a principal connection on $LX$.

A curvature of a linear world connection is defined as that of the
connection (\ref{B}). It reads
\mar{1203}\ben
&& R=\frac12R_{\la\m}{}^\al{}_\bt\dot x^\bt dx^\la\w dx^\m\ot\dot\dr_\al,
\label{1203}\\
&& R_{\la\m}{}^\al{}_\bt = \dr_\la \G_\m{}^\al{}_\bt - \dr_\m
\G_\la{}^\al{}_\bt + \G_\la{}^\g{}_\bt \G_\m{}^\al{}_\g -
\G_\m{}^\g{}_\bt \G_\la{}^\al{}_\g. \nonumber
\een
Due to the canonical splitting of the vertical tangent bundle
\mar{gg4}\beq
VTX=TX\times TX \label{gg4}
\eeq
 of $TX$, the curvature $R$ (\ref{1203}) can be
represented by a tangent-valued two-form
\mar{1203a}\beq
R=\frac12R_{\la\m}{}^\al{}_\bt\dot x^\bt dx^\la\w dx^\m\ot\dr_\al
\label{1203a}
\eeq
on $TX$. Due to this representation, the \textbf{Ricci tensor}
\mar{ric}\beq
R_c=\frac12R_{\la\m}{}^\la{}_\bt dx^\m\ot dx^\bt \label{ric}
\eeq
of a linear world connection $\G$ is defined.

\begin{remark} \label{gg5}
The vertical splitting (\ref{gg4}) with respect to the holonomic
atlases (\ref{tty}) of $TX$ takes place only. Accordingly, the
Ricci tensor (\ref{ric}) with respect to holonomic atlases is ill
defined.
\end{remark}

By a \textbf{torsion} of a linear world connection is meant that
of the connection $\G$ (\ref{B}) on the tangent bundle $TX$ with
respect to the canonical soldering form
\mar{gg11}\beq
\thh_J=dx^\m\ot\dot\dr_\m \label{gg11}
\eeq
on $TX$. It reads
\mar{191}\beq
T =\frac12 T_\m{}^\n{}_\la  dx^\la\w dx^\m\ot \dot\dr_\n, \qquad
T_\m{}^\n{}_\la  = \G_\m{}^\n{}_\la - \G_\la{}^\n{}_\m.
\label{191}
\eeq
A world connection is said to be symmetric if its torsion
(\ref{191}) vanishes, i.e., $\G_\m{}^\n{}_\la = \G_\la{}^\n{}_\m$.
Owing to the vertical splitting of $VTX$, the torsion form
$T$ (\ref{191}) of $\G$ can be written as a tangent-valued
two-form
\mar{mos164}\beq
T =\frac12 T_\m{}^\n{}_\la  dx^\la\w dx^\m\ot \dr_\n
\label{mos164}
\eeq
on $X$.

Being associated to a principal connection on $LX$, a world
connection is represented by a section of the quotient bundle
\mar{015}\beq
C_{\mathrm{W}}=J^1LX/GL_4\to X, \label{015}
\eeq
where $J^1LX$ is the first order jet manifold of sections of
$LX\to X$. We agree to call $C_{\mathrm{W}}$ (\ref{015}) the
bundle of world connections \cite{book09,book00,book13} . With
respect to the holonomic atlas $\Psi_T$ (\ref{tty}), it is
provided with the bundle coordinates $(x^\la, k_\la{}^\nu{}_\al)$
so that, for any section $\G$ of $C_{\mathrm W}\to X$, its
coordinates $k_\la{}^\nu{}_\al\circ \G=\G_\la{}^\nu{}_\al$ are
components of the world connection $\G$ (\ref{B}).

Though the bundle of world connections $C_{\mathrm W}\to X$
(\ref{015}) is not $LX$-associated, it is a natural bundle. It
admits a functorial lift
\be
 \wt\tau_C = \tau^\m\dr_\m
+[\dr_\nu\tau^\al k_\m{}^\nu{}_\bt - \dr_\bt\tau^\nu
k_\m{}^\al{}_\nu - \dr_\m\tau^\nu k_\nu{}^\al{}_\bt +
\dr_{\m\bt}\tau^\al]\frac{\dr}{\dr k_\m{}^\al{}_\bt}
\ee
of any vector field $\tau$ on $X$.

The first order jet manifold $J^1C_{\mathrm W}$ of a bundle of
world connections possesses the canonical splitting
\mar{0101}\ben
&&k_{\la\m}{}^\al{}_\bt =
 \frac12(k_{\la\m}{}^\al{}_\bt - k_{\m\la}{}^\al{}_\bt +
k_\la{}^\g{}_\bt k_\m{}^\al{}_\g -k_\m{}^\g{}_\bt
k_\la{}^\al{}_\g) + \label{0101}\\
&& \qquad \frac12(k_{\la\m}{}^\al{}_\bt +
k_{\m\la}{}^\al{}_\bt - k_\la{}^\g{}_\bt k_\m{}^\al{}_\g +
k_\m{}^\g{}_\bt k_\la{}^\al{}_\g)=\frac12(\cR_{\la\m}{}^\al{}_\bt
+\cS_{\la\m}{}^\al{}_\bt) \nonumber
\een
so that, if $\G$ is a section of $C_{\mathrm W}\to X$, then
$\cR_{\la\m}{}^\al{}_\bt\circ J^1\G=R_{\la\m}{}^\al{}_\bt$ are
components of the curvature (\ref{1203}) \cite{book09,book00}.

\begin{remark} \label{gg2} \mar{gg2}
A world manifold $X$ is called flat if it admits a flat world
connection $\G$, called the Weitzenb\"ock connection. By virtue of
the well-known theorem, there exists a bundle atlas of $TX$ with
constant transition functions such that $\G=dx^\la\ot\dr_\la$
relative to this atlas. However, such an atlas is not holonomic in
general. Therefore, the torsion form $T$ (\ref{191}) of a flat
connection $\G$ need not vanish. A world manifold $X$ is called
parallelizable if the tangent bundle $TX\to X$ is trivial. A
parallelizable manifold is flat. A flat manifold is parallelizable
if it is simply connected. Flat connections together with global
frame fields (Remark \ref{gg3}) on a parallelizable world manifold
are attributes of theory of teleparallel gravity \cite{cai,obu}.
\end{remark}

\section{Lorentz reduced structure}

As was mentioned above, gravitation theory on a world manifold $X$
is classical field theory with spontaneous symmetry breaking
described by Lorentz reduced structures of a frame bundle $LX$
\cite{book09,iva,sardz,sard11}. We deal with the following Lorentz
and proper Lorentz reduced structures.

By a \textbf{Lorentz reduced structure} is meant a reduced
principal $SO(1,3)$-subbundle $L^gX$, called the Lorentz
subbundle, of a frame bundle $LX$. By virtue of the Theorem
\ref{red}, there is one-to-one correspondence between the
principal Lorentz subbundles $L^gX$ of a frame bundle $LX$ and the
global sections of $g$ the quotient bundle $\Si_{\mathrm{PR}}$
(\ref{b3203}) which are \textbf{pseudo-Riemannian metrics} of
signature $(+,---)$ on a world manifold $X$.  For the sake of
convenience, one usually identifies the quotient bundle
$\Si_{\mathrm{PR}}$ (\ref{b3203}), called the \textbf{metric
bundle}, with an open subbundle of the tensor bundle
$\Si_{\mathrm{PR}}\subset \op\vee^2 TX$. Therefore, a metric
bundle $\Si_{\mathrm{PR}}$ can be equipped with bundle coordinates
$(x^\la, \si^{\m\nu})$.

Let L$=SO^0(1,3)$ be a \textbf{proper Lorentz group}, i.e., a
connected component of the unit of $SO(1,3)$. Recall that
$SO(1,3)=\mathbb Z_2\times$L, where $\mathbb Z_2$ is the total
reflection group. A \textbf{proper Lorentz reduced structure} is
defined as a reduced L-subbundle $L^hX$ of $LX$. One needs the
proper Lorentz reduced structure when Dirac spinor fields in
gravitation theory are considered (Section 11).

If a world manifold $X$ is simply connected, there is one-to-ne
correspondence between the Lorentz and proper Lorentz reduced
structures.

One can show that different proper Lorentz subbundles $L^hX$ and
$L^{h'}X$ of a frame bundle $LX$ are isomorphic as principal
L-bundles. This means that there exists a vertical automorphism of
a frame bundle $LX$ which sends $L^hX$ onto $L^{h'}X$. If a world
manifold $X$ is simply connected, the similar property of Lorentz
subbundles also is true.

There is the well-known topological obstruction to the existence
of a Lorentz structure on a world manifold $X$. All non-compact
manifolds and compact manifolds whose Euler characteristic equals
zero admit a Lorentz reduced structure \cite{dods,sardz}.

By virtue of Theorem \ref{red}, there is one-to-one correspondence
between the principal L-subbundles $L^hX$ of a frame bundle $LX$
and the global sections $h$ of the quotient bundle
\mar{5.15}\beq
\Si_{\mathrm T}=LX/\rL\to X,  \label{5.15}
\eeq
called the \textbf{tetrad bundle}. This is an $LX$-associated
bundle with a typical fibre $GL_4/$L. Its global sections are
named the \textbf{tetrad fields}. The fibre bundle (\ref{5.15}) is
a two-fold covering $\zeta: \Si_{\mathrm T}\to \Si_{\mathrm{PR}}$
of the metric bundle $\Si_{\mathrm{PR}}$ (\ref{b3203}). In
particular, every tetrad field $h$ defines a unique
pseudo-Riemannian metric $g=\zeta\circ h$.

Every tetrad field $h$ defines an associated Lorentz bundle atlas
\mar{lat}\beq
\Psi^h=\{(U_\iota,z_\iota^h=\{h_a\})\} \label{lat}
\eeq
of a frame bundle $LX$ such that the corresponding local sections
$z_\iota^h$ of $LX$ take their values into a proper Lorentz
subbundle $L^hX$ and the transition functions of $\Psi^h$
(\ref{lat}) between the frames $\{h_a\}$ are L-valued. The frames
(\ref{lat}):
\mar{b3211a}\beq
\{h_a =h_a^\m(x)\dr_\m\}, \qquad h_a^\m=H_a^\m\circ z_\iota^h,
\qquad x\in U_\iota, \label{b3211a}
\eeq
are called the tetrad frames.

Given a Lorentz bundle atlas $\Psi^h$, the pull-back
\mar{b3211}\beq
h=h^a\ot t_a=z_\iota^{h*}\thh_{LX}=h_\la^a(x) dx^\la\ot t_a
\label{b3211}
\eeq
of the canonical form $\thh_{LX}$ (\ref{b3133'}) by a local
section $z_\iota^h$ is called the (local) tetrad form. It
determines tetrad coframes
\mar{b3211'}\beq
\{h^a =h^a_\m(x)dx^\m\}, \qquad x\in U_\iota, \label{b3211'}
\eeq
in the cotangent bundle $T^*X$. They are the dual of the tetrad
frames (\ref{b3211a}). The coefficients $h_a^\m$ and $h^a_\m$ of
the tetrad frames (\ref{b3211a}) and coframes (\ref{b3211'}) are
called the tetrad functions. They are transition functions between
the holonomic atlas $\Psi_T$ (\ref{tty}) and the Lorentz atlas
$\Psi^h$ (\ref{lat}) of a frame bundle $LX$.

With respect to the Lorentz atlas $\Psi^h$ (\ref{lat}), a tetrad
field $h$ can be represented by the $\mathbb R^4$-valued tetrad
form (\ref{b3211}). Relative to this atlas, the corresponding
pseudo-Riemannian metric $g=\zeta\circ h$ takes the well-known
form
\mar{mos175}\beq
g=\eta(h\ot h)=\eta_{ab}h^a\ot h^b, \qquad
g_{\m\nu}=h_\m^ah_\nu^b\eta_{ab}, \label{mos175}
\eeq
where $\eta=\mathrm{diag}(1,-1,-1,-1)$ is the Minkowski metric in
$\mathbb R^4$ written with respect to its fixed basis $\{t_a\}$.
It is readily observed that the tetrad coframes $\{h^a\}$
(\ref{b3211'}) and the tetrad frames $\{h_a\}$ (\ref{b3211a}) are
orthornormal relative to the pseudo-Riemannian metric
(\ref{mos175}), namely:
\be
g^{\m\nu}h^a_\mu h^b_\nu=\eta^{ab}, \qquad g_{\m\nu}h_a^\mu
h_b^\nu=\eta_{ab}.
\ee
Therefore, their components $h^0$, $h_0$ and $h^i$, $h_i$,
$i=1,2,3$, are called time-like and spatial, respectively.

\begin{remark} \label{gg7} \mar{gg7} It should be emphasized the
difference between tetrad and frame fields. Tetrad fields are
global sections of the quotient bundle $\Si_{\mathrm T}=LX/\rL$
(\ref{5.15}), whereas frame fields are local sections of a frame
bundle $LX$. Since there is one-to-one correspondence between
these sections $h$ and principal L-subbundles $L^hX$ of a frame
bundle $LX$, a tetrad field $h$ locally is represented by a family
of particular frame fields $z^h_i$ (\ref{lat}) taking values into
the corresponding Lorentz subbundle $L^hX\subset LX$, but modulo
L-valued transition functions.
\end{remark}

Given a pseudo-Riemannian metric $g$, any linear world connection
$\G$ (\ref{B}) admits a splitting
\mar{mos191}\beq
\G_{\m\n\al}=\{_{\m\n\al}\} +S_{\m\n\al} +\frac12 C_{\m\n\al}
\label{mos191}
\eeq
in \textbf{Christoffel symbols}
\mar{b1.400}\beq
\{_{\m\n\al}\}= -\frac12(\dr_\m g_{\nu\al} + \dr_\al
g_{\nu\m}-\dr_\nu g_{\m\al}), \label{b1.400}
\eeq
a \textbf{non-metricity tensor}
\mar{mos193}\beq
C_{\m\n\al}=C_{\m\al\n}=\nabla^\G_\m g_{\n\al}=\dr_\m g_{\n\al}
+\G_{\m\n\al} + \G_{\m\al\n}, \label{mos193}
\eeq
and a \textbf{contorsion}
\mar{mos202}\beq
S_{\m\n\al}=-S_{\m\al\n}=\frac12(T_{\n\m\al} +T_{\n\al\m} +
T_{\m\n\al}+ C_{\al\n\m} -C_{\n\al\m}), \label{mos202}
\eeq
where $T_{\m\nu\al}=-T_{\al\nu\m}$ are coefficients of the torsion
form (\ref{mos164}) of $\G$. The tensor fields $T$ and $C$, in
turn, are decomposed into three and four irreducible summands,
respectively \cite{hehl,mcrea}.

A linear world connection $\G$ is called the \textbf{metric
connection} for a pseudo-Riemannian metric $g$ if $g$ is its
integral section, i.e., the metricity condition
\mar{mos203}\beq
\nabla^\G_\m g_{\n\al}=0 \label{mos203}
\eeq
holds. A metric connection reads
\mar{mos204}\beq
\G_{\m\n\al}=\{_{\m\n\al}\} + \frac12(T_{\n\m\al} +T_{\n\al\m} +
T_{\m\n\al}). \label{mos204}
\eeq
The \textbf{Levi--Civita connection}, by definition, is a
torsion-free metric connection $\G_{\m\n\al}=\{_{\m\n\al}\}$.

A principal connection on a proper Lorentz subbundle $L^hX$ of a
frame bundle $LX$ is called the \textbf{Lorentz connection}. Since
connections on a principal bundle are equivariant, this Lorentz
connection is extended to a principal connection $\G$ on a frame
bundle $LX$. The associated linear connection (\ref{B}) on the
tangent bundle $TX$ with respect to the Lorentz atlas $\Psi^h$
(\ref{lat}) reads
\mar{b3205}\beq
\G=dx^\la\ot(\dr_\la + \frac12A_\la{}^{ab}
L_{ab}{}^c{}_dh^d_\m\dot x^\m h_c^\nu\dot\dr_\nu) \label{b3205}
\eeq
where
\be
L_{ab}{}^c{}_d= \eta_{bd}\dl^c_a - \eta_{ad}\dl^c_b
\ee
are generators of a right Lie algebra $\ccG_\rL$ of a proper
Lorentz group L in a Minkowski space $\mathbb R^4$. Written
relative to the holonomic atlas $\Psi_T$ (\ref{tty}), the
connection $\G$ (\ref{b3205}) possesses components
\mar{mos190}\beq
\G_\la{}^\m{}_\nu = h^k_\nu\dr_\la h^\m_k + \eta_{ka}h^\m_b
h^k_\nu A_\la{}^{ab}. \label{mos190}
\eeq
This also is called the Lorentz connection. Its holonomy group is
a subgroup of the proper Lorentz group L. Conversely, let $\G$ be
a world connection with the holonomy group $L$. By virtue of the
well known theorem \cite{kob,book00}, it defines a Lorentz
subbundle of a frame bundle $LX$, and is a Lorentz connection on
this subbundle (see also \cite{sch}).

One can show that any Lorentz connections is a metric world
connection for some pseudo-Riemannian metric $g$ (which is not
necessarily unique \cite{thom}), and {\it vice versa}
\cite{kob,book00},.

At the same time, any linear world connection $\G$ (\ref{B})
yields a Lorentz connection $\G_h$ on each principal L-subbundle
$L^hX$ of a frame bundle \cite{book09,book00,sard11}. It follows
from the fact that the Lie algebra of $GL_4$ is a direct sum
\mar{g13}\beq
\ccG_{GL_4} = \ccG_{\rL} \oplus {\mathfrak m} \label{g13}
\eeq
of the Lie algebra $\ccG_{\rL}$ of a Lorentz group and a subspace
${\mathfrak m}$ such that $[\ccG_{\rL},{\mathfrak m}]\subset
{\mathfrak m}$. Therefore, let us consider a local connection
one-form of a connection $\G$ with respect to the Lorentz atlas
$\Psi^h$ (\ref{lat}) of $LX$ given by tetrad coframes $h^a$
(\ref{b3211'}). It reads
\be
z_\iota^{h*}\ol \G=- \G_\la{}^b{}_a dx^\la\ot L_b{}^a,\qquad
\G_\la{}^b{}_a = -h^b_\m \dr_\la h^\m_a  + \G_\la{}^\m{}_\nu
h^b_\m h^\nu_a,
\ee
where $\{L_b^a\}$ is a basis for a Lie algebra $\ccG_{GL_4}$. The
Lorentz part of this form is precisely a local connection one-form
of a connection $\G_h$ on $L^hX$. We have
\mar{K102}\beq
z_\zeta^{h*}\ol \G_h= -\frac12 A_\la{}^{ab}dx^\la\ot L_{ab},
\qquad A_\la{}^{ab} =\frac12
(\eta^{kb}h^a_\m-\eta^{ka}h^b_\m)(\dr_\la h^\m_k -
 h^\nu_k \G_\la{}^\m{}_\nu). \label{K102}
\eeq
Then combining this expression and the expression (\ref{b3205})
gives a connection
\mar{a3205}\beq
\G_h=dx^\la\ot(\dr_\la +
\frac14(\eta^{kb}h^a_\m-\eta^{ka}h^b_\m)(\dr_\la h^\m_k - h^\nu_k
\G_\la{}^\m{}_\nu) L_{ab}{}^c{}_dh^d_\m\dot x^\m
h_c^\nu\dot\dr_\nu) \label{a3205}
\eeq
with respect to a Lorentz atlas $\Psi^h$ and this connection
\mar{ppp}\beq
\G_h=dx^\la\ot[\dr_\la
+\frac12(h^k_\al\dl^\bt_\m-\eta^{kc}g_{\m\al}h^\bt_c) (\dr_\la
h^\m_k - h^\nu_k \G_\la{}^\m{}_\nu)\dot x^\al\dr_\bt] \label{ppp}
\eeq
relative to a holonomic atlas. If $\G$ is the Lorentz connection
(\ref{mos190}) extended from $L^hX$, then obviously $\G_h=\G$.

\section{Space-time structure}

There is the well-known theorem \cite{book09,higgs,sard14}.

\begin{theorem} \label{gg21} \mar{gg21}
A structure Lie group $G$ of a principal bundle over a paracompact
manifold always is reducible to its maximal compact subgroup $H$.
\end{theorem}

This follows from Theorem \ref{red} and the facts that the
quotient $G/H$ of a Lie group $G$ by its maximal compact subgroup
$H$ is diffeomorphic to an Euclidean space $\mathbb R^m$ and that
a fibre bundle over a paracompact manifold admits a global section
if its typical fibre is an Euclidean space $\mathbb R^m$
\cite{ste}.

A corollary of Theorem \ref{gg21}) is that a structure group
$GL_4$ of a frame bundle $LX$ is reducible to its maximal compact
subgroup $SO(4)$. In gravitation theory, if a structure group
$GL_4$ of $LX$ is reducible to a proper Lorentz group L, it is
always reducible to the maximal compact subgroup $SO(3)$ of L.
Thus, there is a commutative diagram
\mar{8}\beq
\begin{array}{ccc}
 GL_4 &  \longrightarrow & SO(4)  \\
\put(0,10){\vector(0,-1){20}} & & \put(0,10){\vector(0,-1){20}} \\
 \mathrm{L} & \longrightarrow & SO(3)
\end{array} \label{8}
\eeq
of the reduction of structure groups of a frame bundle $LX$ in
gravitation theory. This reduction diagram results in the
following.

$\bullet$ There is one-to-one correspondence between the reduced
principal $SO(4)$-subbundles $L^{g^R}X$ of a frame bundle $LX$ and
the global sections of the quotient bundle $LX/SO(4)\to X$. Its
global sections are Riemannian metrics $g^R$ on $X$. Thus, a
Riemannian metric on a world manifold always exists.

$\bullet$ As was mentioned above, a reduction of a structure group
of a frame bundle $LX$ to a proper Lorentz group implies the
existence of a reduced proper Lorentz subbundle $L^hX\subset LX$
associated to a tetrad field $h$ or a pseudo-Riemannian metric
$g=\zeta\circ h$ on $X$.

$\bullet$ Since a structure group $L$ of this reduced Lorentz
bundle $L^hX$ is reducible to a group $SO(3)$, there exists a
reduced principal $SO(3)$-subbundle
\mar{spat}\beq
L^h_0X\subset L^hX\subset LX, \label{spat}
\eeq
called the spatial structure. The corresponding global section of
the  quotient fibre bundle $L^hX/SO(3)\to X$ with a typical fibre
$\mathbb R^3$ is a one-codimensional spatial distribution
$\bF\subset TX$ on $X$. Its annihilator is a one-dimensional
codistribution $\bF^*\subset T^*X$.

Given the spatial structure $L^h_0X$ (\ref{spat}), let us consider
the Lorentz bundle atlas $\Psi^h_0$ (\ref{lat}) given by local
sections $z_\iota$ of $LX$ taking their values into a reduced
$SO(3)$-subbundle $L^h_0X$. Its transition functions are
$SO(3)$-valued.

It follows that, in gravitation theory on a world manifold $X$,
one can always choose an atlas of the tangent bundle $TX$ and
associated bundles with $SO(3)$-valued transition functions. It is
called the spatial bundle atlas.

Given a spatial bundle atlas $\Psi^h_0$, its $SO(3)$-valued
transition functions preserve a time-like component
\mar{h0a}\beq
h^0=h^0_\la dx^\la \label{h0a}
\eeq
of local tetrad forms (\ref{b3211}) which, therefore, is globally
defined. We agree to call it the time-like tetrad form.
Accordingly, the dual time-like vector field
\mar{h0b}\beq
h_0=h^\m_0\dr_\m \label{h0b}
\eeq
also is globally defined. In this case, a spatial distribution
$\bF$ is spanned by spatial components $h_i$, $i=1,2,3$, of the
tetrad frames (\ref{b3211a}), while the time-like tetrad form
(\ref{h0a}) spans the tetrad codistribution $\bF^*$, i.e.,
\mar{yuy2}\beq
h^0\rfloor \bF=0. \label{yuy2}
\eeq
Then the tangent bundle $TX$ of a world manifold $X$ admits a
space-time decomposition
\mar{mos209}\beq
TX=\bF\oplus T^0X, \label{mos209}
\eeq
where $T^0X$ is a one-dimensional fibre bundle spanned by the
time-like vector field $h_0$ (\ref{h0b}).

Due to the commutative diagram (\ref{8}), the reduced L-subbundle
$L^h_0X$ (\ref{spat}) of a reduced Lorentz bundle $L^hX$ is a
reduced subbundle of some reduced $SO(4)$-bundle $L^{g^R}X$ too,
i.e.,
\mar{yuy}\beq
L^hX\supset L^h_0X\subset L^{g^R}X. \label{yuy}
\eeq
Let $g=\zeta\circ h$ and $g^R$ be the corresponding
pseudo-Riemannian and Riemannian metrics on $X$. Written with
respect to a spatial bundle atlas $\Psi^h_0$, they read
\mar{prm1,2}\ben
&& g=\eta_{ab}h^a\ot h^b, \qquad
g_{\m\nu}=h_\m^ah_\nu^b\eta^{ab}, \label{prm1}\\
&& g^R=\eta^E_{ab}h^a\ot h^b, \qquad
g^R_{\m\nu}=h_\m^ah_\nu^b\eta^E_{ab}, \label{prm2}
\een
where $\eta^E$ is an Euclidean metric in $\mathbb R^4$. The
space-time decomposition (\ref{mos209}) is orthonormal with
respect to both the metrics (\ref{prm1}) and (\ref{prm2}). Thus,
we come to the following well-known results
\cite{book09,haw,sardz}.

$\bullet$ For any pseudo-Riemannian metric $g$ on a world manifold
$X$, there exist a normalized time-like one-form $h^0$ and a
Riemannian metric $g^R$ such that
\mar{mos208}\beq
g=2h^0\ot h^0 -g^R. \label{mos208}
\eeq
Conversely, let a world manifold $X$ admit a nowhere vanishing
one-form $\si$ (or, equivalently, a nowhere vanishing vector
field). Then  any Riemannian world metric $g^R$ on $X$ yields the
pseudo-Riemannian world metric $g$ (\ref{mos208}) where
$h^0=\si(g^R(\si,\si))^{-1/2}$.

$\bullet$ A world manifold $X$ admits a pseudo-Riemannian metric
iff there exists a nowhere vanishing one-form (or a vector field)
on $X$.

Note that the condition (\ref{yuy}) gives something more. Namely,
there is  one-to-one correspondence between the reduced
$SO(3)$-subbundles of a frame bundle $LX$ and the triples
$(g,\bF,g^R)$  of  a pseudo-Riemannian metric $g$, a spatial
distribution $\bF$ defined by the condition (\ref{yuy2}) and a
Riemannian metric $g^R$ which obey the relation (\ref{mos208}). A
spatial distribution $\bF$ and a Riemannian metric $g^R$ in the
triple $(g,\bF,g^R)$ are called $g$-compatible. The corresponding
space-time decomposition is said to be a $g$-compatible
\textbf{space-time structure}. A world manifold endowed with a
pseudo-Riemannian metric and a compatible space-time structure is
called the \textbf{space-time}.

\begin{remark}
A $g$-compatible Riemannian metric  $g^R$   in  a triple
$(g,\bF,g^R)$ defines a $g$-compatible distance function $d(x,x')$
on a world manifold $X$.  Such a function brings $X$ into a metric
space whose locally Euclidean topology is equivalent to a manifold
topology on $X$. Given a gravitational field $g$, the
$g$-compatible Riemannian metrics and the corresponding distance
functions are different for different spatial distributions $\bF$
and $\bF'$.  It  follows that physical observers associated  to
different spatial distributions $\bF$ and $\bF'$ perceive a world
manifold $X$ as different Riemannian spaces. The well-known
relativistic changes of sizes of moving bodies exemplify this
phenomenon. Note that there were attempts of deriving a world
topology directly from its pseudo-Riemannian structure (e.g., path
topology, $C^0$-topology, etc.) \cite{gob,haw}. However, these
topologies are rather extraordinary, e.g., they are the
non-Hausdorff ones.
\end{remark}

\section{Metric-affine gauge gravitation theory}

In the absence of matter fields, dynamic variables of gauge
gravitation theory are linear world connections and
pseudo-Riemannian metrics on $X$ \cite{book09,book00,higgs3}.
Their Lagrangian $L_{\mathrm{MA}}$ is invariant under general
covariant transformations.

This is the case of metric-affine gravitation theory
\cite{blag,cap,hehl,heh07,mcrea,obukh}. Let us however emphasize
that we consider general linear connections which need not be
metric (Lorentz) connections.

\begin{remark} In view of the decomposition (\ref{mos191}), one can
choose a different collection of dynamic variables of
metric-affine gauge gravitation theory. These are a
pseudo-Riemannian metric, the torsion (\ref{191}) and the
non-metricity tensor (\ref{mos193}).
\end{remark}

World connections are represented by sections of the bundle of
world connections $C_{\mathrm W}$ (\ref{015}). World metrics are
described by sections of the quotient bundle (\ref{b3203}).
Therefore, let us consider the bundle product
\mar{gg9}\beq
Y_{\mathrm MA}=\Si_{\mathrm{PR}}\op\times_X C_{\mathrm W},
\label{gg9}
\eeq
coordinated by $(x^\la,\si^{\m\nu}, k_\mu{}^\al{}_\bt)$.

Let us restrict our consideration to first order Lagrangian theory
on $Y_{\mathrm MA}$. Then a configuration space of gauge
gravitation theory is the first order jet manifold
\mar{kkl}\beq
J^1Y_{\mathrm MA}= J^1\Si_{\mathrm{PR}}\op\times_X J^1C_{\mathrm
W}, \label{kkl}
\eeq
coordinated by $(x^\la,\si^{\m\nu}, k_\mu{}^\al{}_\bt,
\si_\la^{\m\nu}, k_{\la\mu}{}^\al{}_\bt)$
\cite{book09,sard11,book13}. A first order Lagrangian
$L_{\mathrm{MA}}$ of metric-affine gauge gravitation theory is a
defined as a density
\mar{10130}\beq
L_{\mathrm{MA}}=\cL_{\mathrm{AM}}(x^\la,\si^{\m\nu},
k_\mu{}^\al{}_\bt, \si_\la^{\m\nu}, k_{\la\mu}{}^\al{}_\bt)\om,
\qquad \om=dx^1\w\cdots\w dx^4, \label{10130}
\eeq
on the configuration space $J^1Y$ (\ref{kkl}). Its Euler--Lagrange
operator is
\be
&& \dl L_{\mathrm{MA}}= (\cE_{\al\bt} d\si^{\al\bt} + \cE^\m{}_\al{}^\bt
dk_\m{}^\al{}_\bt)\w \om.  \\
&& \cE_{\al\bt} =\left(\frac{\dr}{\dr \si^{\al\bt}}
 - d_\la\frac{\dr}{\dr \si^{\al\bt}_\la}\right)\cL_{\mathrm{AM}},\qquad
\cE^\m{}_\al{}^\bt =\left(\frac{\dr}{\dr k_\mu{}^\al{}_\bt}
 - d_\la\frac{\dr}{\dr k_{\la\mu}{}^\al{}_\bt}\right)\cL_{\mathrm{AM}}, \\
&& d_\la=\dr_\la + \si^{\al\bt}_\la \frac{\dr}{\dr \si^{\al\bt}} +
k_{\la\mu}{}^\al{}_\bt \frac{\dr}{\dr k_\mu{}^\al{}_\bt} +
\si^{\al\bt}_{\la\nu} \frac{\dr}{\dr \si^{\al\bt}_\nu} +
k_{\la\nu\mu}{}^\al{}_\bt \frac{\dr}{\dr k_{\nu\mu}{}^\al{}_\bt}.
\ee
The corresponding Euler--Lagrange equations read
\be
\cE_{\al\bt}=0, \qquad  \cE^\m{}_\al{}^\bt=0.
\ee

The fibre bundle $Y_{\mathrm MA}$ (\ref{gg9}) is a natural bundle
admitting the functorial lift
\mar{gr3}\ben
&& \wt\tau_{\Si C}=\tau^\m\dr_\m +(\si^{\nu\bt}\dr_\nu \tau^\al
+\si^{\al\nu}\dr_\nu \tau^\bt)\frac{\dr}{\dr \si^{\al\bt}} +
\label{gr3}\\
&& \qquad (\dr_\nu \tau^\al k_\m{}^\nu{}_\bt -\dr_\bt \tau^\nu
k_\m{}^\al{}_\nu -\dr_\mu \tau^\nu k_\nu{}^\al{}_\bt
+\dr_{\m\bt}\tau^\al)\frac{\dr}{\dr k_\mu{}^\al{}_\bt} \nonumber
\een
of vector fields $\tau$ on $X$. It is an infinitesimal generator
of general covariant transformations. At the same time,
$\wt\tau_{\Si C}$ (\ref{gr3}) also is a gauge transformation whose
gauge parameters are components $\tau^\la(x)$ of vector fields
$\tau$ on $X$.

By virtue of Relativity Principle, the Lagrangian
$L_{\mathrm{MA}}$ (\ref{10130}) of metric-affine gauge gravitation
theory is assumed to be invariant under general covariant
transformations. Its Lie derivative along the jet prolongation
$J^1\wt\tau_{\Si C}$ of the vector field $\wt\tau_{\Si C}$
(\ref{gr3}) for any $\tau$ vanishes, i.e.,
\mar{10140}\beq
\bL_{J^1\wt\tau_{\Si C}}L_{\mathrm{MA}}=0. \label{10140}
\eeq

Since a configuration space $J^1C_{\mathrm W}$ of world
connections possesses the canonical splitting (\ref{0101}), the
following analogy to the well-known Utiyama theorem in Yang--Mills
gauge theory is true.

\begin{theorem} \label{httu1} \mar{httu1} If the first order Lagrangian $L_{\mathrm{MA}}$ (\ref{10130})
on the configuration space (\ref{kkl}) is invariant under general
covariant transformations and it does not depend on the jet
coordinates $\si^{\al\bt}_\la$ (i.e., derivatives of a metric),
this Lagrangian factorizes through the terms
$\cR_{\la\m}{}^\al{}_\bt$ (\ref{0101}).
\end{theorem}

In contrast with the well-known Lagrangian of Yang--Mills gauge
theory, different contractions of a curvature tensor
$\cR_{\la\m}{}^\al{}_\bt$ are possible. For instance, the Ricci
tensor $R_c$ (\ref{ric}) and a scalar curvature $\cR$ are defined.
Moreover, a Lagrangian $L_{\mathrm{MA}}$ also can depend
separately on a torsion
\mar{10145}\beq
t_\m{}^\nu{}_\la = k_\m{}^\nu{}_\la - k_\la{}^\nu{}_\m.
\label{10145}
\eeq

\begin{example} \label{hhtuu} \mar{hhtuu}
In metric-affine gravitation theory, the Hilbert--Einstein
Lagrangian of General Relativity takes a form
\mar{10221}\beq
L_\mathrm{GR}=\cR\sqrt{\si}\om =\si^{\m\bt}
\cR_{\la\m}{}^\la{}_{\bt}\sqrt{\si}\om. \label{10221}
\eeq
The corresponding Euler--Lagrange equations read
\mar{10180,1}\ben
&& \cE_{\al\bt}=\cR_{\al\bt} -\frac12 \si_{\al\bt} \cR=0, \label{10180}\\
&& \cE^\nu{}_\al{}^\bt=-d_\al(\si^{\nu\bt} \sqrt{\si})
+d_\la(\si^{\la\bt}\sqrt{\si})\dl^\nu_\al + \label{10181}\\
&& \qquad (\si^{\nu\g} k_\al{}^\bt{}_\g -\si^{\la\g}\dl^\nu_\al
k_\la{}^\bt{}_\g - \si^{\nu\bt}k_\g{}^\g{}_\al
+\si^{\la\bt}k_\la{}^\nu{}_\al)\sqrt{\si}=0. \nonumber
\een
The equation (\ref{10180}) is an analogy of the Einstein
equations, whereas the equation (\ref{10181}) describes the
torsion (\ref{10145}) and the non-metricity
\be
c_{\m\nu\al}=c_{\m\al\nu}=d_\m \si_{\nu\al}
+k_\m{}^\bt{}_\al\si_{\nu\bt} + k_\m{}^\bt{}_\nu\si_{\bt\al}
\ee
of a linear world connection. It is brought into a form
\be
&& \sqrt{\si^{-1}}\si_{\nu\ve}\si_{\bt\m}\cE^\nu{}_\al{}^\bt=
 c_{\al\ve\m}-\frac12 \si_{\m\ve}\si^{\la\g}c_{\al\la\g} -
\si_{\al\ve}\si^{\la\bt}c_{\la\bt\m} +  \\
&& \qquad \frac12
\si_{\al\ve}\si^{\la\g}c_{\m\la\g} +
 t_{\m\ve\al} + \si_{\m\ve} t_\al{}^\g{}_\g + \si_{\al\ve} t_\g{}^\g{}_\m
=0.
\ee
\end{example}

\begin{example} \label{gg10} \mar{gg10}
The Yang--Mills Lagrangian
\be
L_\mathrm{YM}=\si^{\m\la}\si^{\nu\g}\cR_{\m\nu}{}^\al{}_\bt
\cR_{\la\g}{}^\bt{}_\al\sqrt{\si}\om
\ee
in metric-affine gauge gravitation theory also is considered. It
is invariant under a total group Aut$(LX)$ of automorphisms of a
frame bundle $LX$ (Remark \ref{gg3}). In this case, metric
variables $\si^{\m\la}$ fail to be dynamic because they are
brought into a constant Minkowski metric by general frame
transformations.
\end{example}

\section{Energy-momentum conservation law}

Since infinitesimal general covariant transformations
$\wt\tau_{\Si C}$ (\ref{gr3}) are exact symmetries of a
metric-affine gravitation Lagrangian, let us study the
corresponding conservation law. This is the energy-momentum
conservation laws because vector fields $\wt\tau_{\Si C}$ are not
vertical \cite{book09,sard97}.  Moreover, since infinitesimal
general covariant transformations $\wt\tau_{\Si C}$ (\ref{gr3})
are gauge transformations depending on derivatives of gauge
parameters, the corresponding energy-momentum current reduces to a
superpotential \cite{book09,sard09}.

In view of Theorem \ref{httu1}, let us assume that the
metric-affine gravitation Lagrangian $L_{\mathrm{MA}}$
(\ref{10130}) is independent of the derivative coordinates
$\si_\la{}^{\al\bt}$ of a world metric and that it factorizes
through the curvature terms $\cR_{\la\m}{}^\al{}_\bt$
(\ref{0101}). Then the following relations hold:
\be
&&  \pi^{\la\nu}{}_\al{}^\bt= -\pi^{\nu\la}{}_\al{}^\bt, \qquad
\pi^{\la\nu}{}_\al{}^\bt=\frac{\dr \cL_{\mathrm{MA}}}{\dr k_{\la\nu}{}^\al{}_\bt}, \\
&&\frac{\dr\cL_{\mathrm{MA}}}{\dr k_\nu{}^\al{}_\bt}=
\pi^{\la\nu}{}_\al{}^\si k_\la{}^\bt{}_\si
-\pi^{\la\nu}{}_\si{}^\bt k_\la{}^\si{}_\al.
\ee

Let us use the compact notation
\be
y^A=k_\m{}^\al{}_\bt, \qquad u_\m{}^\al{}_\bt{}^{\ve\si}_\g =
\dl^\ve_\m \dl^\si_\bt \dl^\al_\g, \qquad
u_\m{}^\al{}_\bt{}^\ve_\g= k_\m{}^\ve{}_\bt \dl^\al_\g
-k_\m{}^\al{}_\g \dl^\ve_\bt - k_\g{}^\al{}_\bt \dl^\ve_\m.
\ee
Then the vector field (\ref{gr3}) takes a form
\be
\wt\tau_{\Si C} =\tau^\la\dr_\la  + (\si^{\nu\bt}\dr_\nu\tau^\al
+\si^{\al\nu}\dr_\nu\tau^\bt)\dr_{\al\bt}+
(u^A{}_\al^\bt\dr_\bt\tau^\al
+u^A{}_\al^{\bt\m}\dr_{\bt\m}\tau^\al)\dr_A.
\ee

Let $L_{\mathrm{MA}}$ be invariant under general covariant
transformations, i.e., the equality (\ref{10140}) for any vector
field $\tau$ is satisfied. On-shell, we then have a weak
conservation law
\mar{K8}\ben
&& 0\ap - d_\la[ \pi^\la_A(y^A_\al\tau^\al -u^A{}_\al^\bt\dr_\bt\tau^\al
 -u^A{}_\al^{\ve\bt}\dr_{\ve\bt}\tau^\al) -\tau^\la\cL_{\mathrm{MA}}] \label{K8}
\een
of the \textbf{energy-momentum current} of metric-affine gravity
\mar{b3190}\beq
\cJ_{\mathrm{MA}}{}^\la= \pi^\la_A(y^A_\al\tau^\al
-u^A{}_\al^\bt\dr_\bt\tau^\al
 -u^A{}_\al^{\ve\bt}\dr_{\ve\bt}\tau^\al)-\tau^\la\cL_{\mathrm{MA}}. \label{b3190}
\eeq

\begin{remark}
It is readily observed that, with respect to a local coordinate
system where a vector field $\tau$ is constant, the
energy-momentum current (\ref{b3190}) leads to a canonical
energy-momentum tensor
\be
\cJ_{\rm MA}{}^\la{}_\al\tau^\al =(\pi^{\la\m}{}_\bt{}^\nu
k_{\al\m}{}^\bt{}_\nu -\dl^\la_\al \cL_{\rm MA})\tau^\al,
\ee
suggested in order to describe an energy-momentum complex in the
Palatini model \cite{dick}.
\end{remark}

Due to the arbitrariness of $\tau^\la$, we have a set of
equalities
\mar{b3173b}\ben
&& \pi^{(\la\ve}{}_\g{}^{\si)}=0, \nonumber\\
&& (u^A{}_\g^{\ve\si}\dr_A + u^A{}_\g^\ve\dr^\si_A)\cL_{\mathrm{MA}}= 0, \nonumber\\
&& \dl^\bt_\al\cL_{\mathrm{MA}} + 2\si^{\bt\m}\dl_{\al\m}\cL_{\mathrm{MA}} + u^A{}_\al^\bt\dl_A\cL_{\mathrm{MA}}
 + d_\m(\pi^\m_A  u^A{}_\al^\bt)
-y^A_\al\pi^\bt_A  =0,\label{b3173b} \\
&& \dr_\la\cL_{\mathrm{MA}}=0. \nonumber
\een
Substituting the term $y^A_\al\pi^\bt_A$ from the expression
(\ref{b3173b}) in the energy-momentum conservation law (\ref{K8}),
one brings this conservation law into a form
\mar{b3174}\ben
&& 0\ap -
d_\la[2\si^{\la\m}\tau^\al\dl_{\al\m}\cL_{\mathrm{MA}} +
u^A{}_\al^\la\tau^\al\dl_A\cL_{\mathrm{MA}} - \pi^\la_Au^A{}_\al^\bt\dr_\bt\tau^\al + \label{b3174}\\
&& \qquad d_\m(\pi^{\la\m}{}_\al{}^\bt)
\dr_\bt\tau^\al + d_\m(\pi^\m_A  u^A{}_\al^\la)\tau^\al -
d_\m(\pi^{\la\m}{}_\al{}^\bt \dr_\bt\tau^\al)]. \nonumber
\een
After separating the variational derivatives, the energy-momentum
conservation law (\ref{b3174}) of a metric-affine gravity takes a
superpotential form
\be
&& 0\ap - d_\la [2\si^{\la\m}\tau^\al\dl_{\al\m}\cL_{\mathrm{MA}}
+(k_\m{}^\la{}_\g\dl^\m{}_\al{}^\g\cL_{\mathrm{MA}} -
 k_\m{}^\si{}_\al\dl^\m{}_\si{}^\la\cL_{\mathrm{MA}} -
k_\al{}^\si{}_\g\dl^\la{}_\si{}^\g\cL_{\mathrm{MA}})\tau^\al +  \\
&& \qquad \dl^\la{}_\al{}^\m\cL_{\mathrm{MA}}\dr_\m\tau^\al
-d_\m(\dl^\m{}_\al{}^\la\cL_{\mathrm{MA}})\tau^\al +
 d_\m(\pi^{\m\la}{}_\al{}^\nu(\dr_\nu\tau^\al
-k_\si{}^\al{}_\nu\tau^\si))],
\ee
where an energy-momentum current on-shell reduces to a
\textbf{generalized Komar superpotential}
\mar{K3}\beq
 U_{\mathrm{MA}}{}^{\m\la}= 2\frac{\dr\cL_{\mathrm{MA}}}{\dr
\cR_{\m\la}{}^\al{}_\nu}(D_\nu\tau^\al +
t_\nu{}^\al{}_\si\tau^\si), \label{K3}
\eeq
where $D_\nu$ is a covariant derivative relative to a connection
$k_\nu{}^\al{}_\si$ and $t_\nu{}^\al{}_\si$ is its torsion
\cite{giacqg,sard97b,sard11}.

In particular, the Hilbert--Einstein Lagrangian (\ref{10221}) is
invariant under general covariant transformations. The
corresponding generalized Komar superpotential (\ref{K3}) comes to
the well-known Komar superpotential if one substitutes the
Levi--Civita connection $k_\nu{}^\al{}_\si =\{_\nu{}^\al{}_\si\}$.

\section{Spinor structure}

In classical field theory, Dirac spinor fields usually are
represented by sections of a spinor bundle on a world manifold $X$
whose typical fibre is a Dirac spinor space $\Psi(1,3)$ and whose
structure group is a Lorentz spin group Spin$(1,3)$. In order to
introduce the Dirac operator, one however must assume that Dirac
spinors carry out a representation of a Clifford algebra.
Moreover, we describe spinor spaces as subspaces of Clifford
algebras and define spinor bundles as subbundles of fibre bundles
in Clifford algebras \cite{book09,sard11,cliff15}.

Note that spinor representations of Lie algebras $so(m,n-m)$ of
pseudo-orthogonal Lie groups $SO(m,n-m)$, $n\geq 1$,
$m=0,1,\ldots,n$, were discovered by E. Cartan in 1913, when he
classified finite-dimensional representations of simple Lie
algebras \cite{cartan}. Though, there is a problem of spinor
representations of pseudo-orthogonal Lie groups $SO(m,n-m)$
themselves. Spinor representations are attributes of Spin groups
Spin$(m,n-m)$. Spin groups Spin$(m,n-m)$ are two-fold coverings
(\ref{104}) of pseudo-orthogonal groups $SO(m,n-m)$.

Spin groups Spin$(m,n-m)$ are defined as certain subgroups of real
Clifford algebras $\cC(m,n-m)$ (\ref{104a}). Moreover, spinor
representations of Spin groups in fact are the restriction of
spinor representation of real Clifford algebras to its Spin
subgroups. As was mentioned above, one needs an action of a whole
real Clifford algebra in a spinor space in order to construct a
Dirac operator. In 1935, R. Brauer and H. Weyl described spinor
representations in terms of Clifford algebras \cite{brauer,law}.
This description is based on the following.

$\bullet$ Real Clifford algebras $\cC(m,n-m)$ and complex Clifford
algebras $\mathbb C\cC(n)$ of even dimension $n$ are isomorphic to
matrix algebras (Theorems \ref{k33} and \ref{k8}, respectively).
Therefore, they are simple, and all their automorphisms are inner
(Theorems \ref{k15} and \ref{k21}). Their invertible elements
constitute general linear matrix groups. They act on Clifford
algebras by a left-regular representation, and their adjoint
representation exhaust all automorphisms of Clifford algebras.

$\bullet$ Given a real Clifford algebra $\cC(m,n-m)$, the
corresponding spinor space $\Psi(m,n-m)$ is defined as a carrier
space of its exact irreducible representation. This representation
of a real Clifford algebra $\cC(m,n-m)$ of even dimension $n$ is
unique up to an equivalence (Theorem \ref{a10}).

However, spinor spaces $\Psi(m,n-m)$ and $\Psi(m',n-m')$ need not
be isomorphic vector spaces for $m'\neq m$. For instance, a Dirac
spinor space is defined to be a spinor space $\Psi(1,3)$ of a real
Clifford algebra $\cC(1,3)$. It differs from a Majorana spinor
space $\Psi(3,1)$ of a real Clifford algebra $\cC(3,1)$. In
contrast with the four-dimensional real matrix representation
(\ref{62}) of $\cC(3,1)$, the representation (\ref{41'}) of a real
Clifford algebra $\cC(3,1)$ by complex Dirac's matrices is not a
representation of a real Clifford algebra. By this reason and
because, from the physical viewpoint, Dirac spinor fields
describing charged fermions are complex fields, we focus our
consideration on complex Clifford algebras and
 complex spinors.

 $\bullet$  A complex Clifford algebra $\mathbb C\cC(n)$ of even dimension $n$
is proved to be isomorphic to a ring $\mathrm{Mat}(2^{n/2},
\mathbb C)$ of complex $(2^{n/2}\times 2^{n/2})$-matrices (Theorem
\ref{k8}). The corresponding complex spinor space $\Psi(n)$ is
defined as a carrier space of its exact irreducible
representation. Due to the canonical monomorphism $\cC(m,n-m)\to
\mathbb C\cC(n)$ (\ref{sp200}) of real Clifford algebras to the
complex ones, a complex spinor space $\Psi(n)$ admits a
representation of a real Clifford algebra $\cC(m,n-m)$, though it
need not be irreducible.

$\bullet$ Similarly to a case of real Clifford algebras, an exact
irreducible representation of a complex Clifford algebra $\mathbb
C\cC(n)$ of even dimension $n$ is unique up to an equivalence
(Theorem \ref{a11}). Therefore, we define a complex spinor space
$\Psi(n)$ in a case of even $n$ as a minimal left ideal of a
complex Clifford algebra $\mathbb C\cC(n)$. Thus, a spinor
representation
\mar{sp110}\beq
\gamma:\mathbb C\cC(n)\times \Psi(n) \to \Psi(n) \label{sp110}
\eeq
of a Clifford algebra $\mathbb C\cC(n)$ is equivalent to the
canonical representation of $\mathrm{Mat}(2^{n/2}, \mathbb C)$ by
matrices in a complex vector space $\Psi(n)=\mathbb C^{2^{n/2}}$.

Treating a complex spinor space $\Psi(n)$ as a subspace of a
complex Clifford algebra $\mathbb C\cC(n)$ which carries out its
left-regular representation (\ref{sp110}), we believe reasonable
to consider a fibre bundle in spinor spaces $\Psi(n)$ as a
subbundle of a fibre bundle in Clifford algebras. However, one
usually considers fibre bundles in Clifford algebras whose
structure group is a group of automorphisms of these algebras
\cite{book09,law}. A problem is that this group fails to preserve
spinor subspaces $\Psi(n)$ of a complex Clifford algebra $\mathbb
C\cC(n)$ (Remark \ref{gg15}) and, thus, it can not be a structure
group of spinor bundles.

Therefore, we define fibre bundles $\lC$ (\ref{sp100}) in Clifford
algebras $\mathbb C\cC(n)$ whose structure group is a general
linear group $GL(2^{n/2}, \mathbb C)$ of invertible elements of
$\mathbb C\cC(n)$ which acts on this algebra by left
multiplications \cite{cliff15}. Certainly, it preserves minimal
left ideals of this algebra and, consequently, is a structure
group of spinor subbundles $S$ of a Clifford algebra bundle $\lC$.

It should be emphasized that, though there is the ring
monomorphism $\cC(m,n-m)\to \mathbb C\cC(n)$ (\ref{sp200}), the
Clifford algebra bundle $\lC$ (\ref{sp100}) need not contain a
subbundle in real Clifford algebras $\cC(m,n-m)$ unless a
structure group $GL(2^{n/2}, \mathbb C)$ of $\lC$ is reducible to
a group $\cG\cC(m,n-m)$ of invertible elements of $\cC(m,n-m)$.
Let $X$ be an $n$-dimensional smooth manifold and $LX$ a principal
frame bundle over $X$. In accordance with Theorem \ref{red}, any
global section $h$ of the quotient bundle
$\Si(m,n-m)=LX/O(m,n-m)\to X$ (\ref{sp226}) is associated to the
fibre bundle $\lC^h\to X$ (\ref{sp240}) in complex Clifford
algebras $\mathbb C\cC(n)$ which contains the subbundle
$\lC^h(m,n-m)\to X$ (\ref{sp241}) in real Clifford algebras
$\cC(m,n-m)$ and a spinor subbundle $S^h\to X$.

A key point is that, given different sections $h$ and $h'$ of the
quotient bundle $\Si(m,n-m)\to X$ (\ref{sp226}), the Clifford
algebra bundles $\lC^h$ and $\lC^{h'}$ need not be isomorphic.

In order to describe all these non-isomorphic Clifford algebra
bundles $\lC^h$, we follow a construction of composite bundles. We
consider composite Clifford algebra bundles $\lC_\Si$
(\ref{sp246}) and $\lC(m,n-m)_\Si$ (\ref{sp247}), and the spinor
bundle $S_\Si$ (\ref{sp248}) over a base $\Si(m,n-m)$
(\ref{sp226}). Then given a global section $h$ of the quotient
bundle $\Si(m,n-m)\to X$ (\ref{sp226}), the pull-back bundles
$h^*\lC_\Si$, $h^*\lC(m,n-m)_\Si$ and $h^*S_\Si$ are the above
mentioned fibre bundles $\lC^h\to X$, $\lC^h(m,n-m)\to X$ and
$S^h\to X$, respectively.

\subsection{Clifford algebras}

A real Clifford algebra is defined as a ring (i.e., a unital
associative algebra) possessing a certain vector subspace of
generating elements. However, such a ring can possess different
generating spaces. Therefore, we also consider a real Clifford
algebra without specifying its generating space.

Let $V=\mathbb R^n$ be an $n$-dimensional real vector space
provided with a non-degenerate bilinear form (a pseudo-Euclidean
metric) $\eta$. Let us consider a tensor algebra
\be
\otimes V= \mathbb{R} \oplus V\oplus \op\otimes^2V\oplus\cdots
\oplus \op\otimes^k V\oplus \cdots
\ee
of $V$ and its two-sided ideal $I_\eta$ generated by the elements
\be
v\otimes v'+v'\otimes v - 2\eta(v,v')e, \qquad  v,v'\in V,
\ee
where $e$ denotes the unit element of $\otimes V$. The quotient
$\otimes V/I_\eta$ is a real non-commutative ring. A real ring
$\otimes V/I_\eta$ together with a fixed generating subspace
$(V,\eta)$ is called the \textbf{real Clifford algebra}
$\cC(V,\eta)$ modelled over a pseudo-Euclidean space $(V,\eta)$.

There is the canonical monomorphism of a real vector space $V$ to
the quotient  $\otimes V/I_\eta$. It is a generating subspace of a
real ring $\otimes V/I_\eta$. Its elements obey the relations
\be
vv'+v'v - 2\eta(v,v')e=0, \qquad  v,v'\in V.
\ee

Given real Clifford algebras $\cC(V,\eta)$ and $\cC(V',\eta')$, by
their isomorphism is meant an isomorphism of them as rings:
\mar{ss2}\beq
\phi:\cC(V,\eta) \to \cC(V',\eta'), \qquad
\phi(qq')=\phi(q)\phi(q'), \label{ss2}
\eeq
which also is an isometric isomorphism of their generating
pseudo-Euclidean spaces:
\mar{ss1}\ben
&& \phi:\cC(V,\eta)\supset (V,\eta)\to
(V',\eta')\subset\cC(V',\eta'), \label{ss1} \\
&& 2\eta'(\phi(v),\phi(v'))=\phi(v)\phi(v') + \phi(v')\phi(v)=\phi(vv' + v'v)
=2\eta(v,v').\nonumber
\een

It follows from the isomorphism (\ref{ss1}) that two real Clifford
algebras $\cC(V,\eta)$ and $\cC(V',\eta')$ are isomorphic iff they
are modelled over pseudo-Euclidean spaces $(V,\eta)$ and
$(V',\eta')$ of the same signature. Let a pseudo-Euclidean metric
$\eta$ be of signature $(m;n-m)=(1,...,1;-1,...,-1)$. Let
$\{v^1,...,v^n\}$ be a basis for $V$ such that $\eta$ takes a
diagonal form
\be
\eta^{ab}=\eta(v^a,v^b)=\pm \delta^{ab}.
\ee
Then a ring $\cC(V,\eta)$ is generated by elements $v^1,...,v^n$
which obey relations
\be
v^a v^b+v^b v^a=2\eta^{ab}e.
\ee
We agree to call $\{v^1,...,v^n\}$ the basis for a real Clifford
algebra $\cC(\mathbb R^n,\eta)$. Given this basis, let us denote
$\cC(\mathbb R^n,\eta)=\cC(m,n-m)$.

Certainly, any isomorphism (\ref{ss2}) -- (\ref{ss1}) of real
Clifford algebras is their ring isomorphism (\ref{ss2}). However,
the converse is not true, because their ring isomorphism
(\ref{ss2}) need not be the isometric isomorphism (\ref{ss1}) of
their generating spaces. Therefore, we also consider real Clifford
algebras, without specifying their generating spaces.

\begin{lemma} \label{ss4} \mar{ss4}
Any isometric isomorphism (\ref{ss1}) of generating vector spaces
$\phi: V\to V'$ of real Clifford algebras $\cC(V,\eta)$ and
$\cC(V',\eta')$ is prolonged to their ring isomorphism
(\ref{ss2}):
\mar{ss3}\beq
\phi:\cC(V,\eta)\to \cC(V',\eta') \qquad \phi(v_1\cdots v_k)=
\phi(v_1)\cdots \phi(v_k), \label{ss3}
\eeq
which also is an isomorphism of real Clifford algebras.
\end{lemma}

\begin{remark}
It may happen that a ring $\cC(V,\eta)$ admits a generating
pseudo-Euclidean space $(V',\eta')$ whose signature differs from
that of $(V,\eta)$. In this case, $\cC(V,\eta)$ possesses the
structure of a real Clifford algebra $\cC(V',\eta')$ which is not
isomorphic to a real Clifford algebra $\cC(V,\eta)$.
\end{remark}

There is the following classification of real Clifford algebras
\cite{law,cliff15}.

\begin{theorem} \label{k33} \mar{k33} Real Clifford algebras $\cC(p,q)$
as rings are isomorphic to the following matrix algebras.
\mar{k5}\beq
\cC(p,q) =  \left\{
\begin{array}{ll}
\mathrm{Mat}(2^{(p+q)/2},\mathbb R)=\op\ot^{(p+q)/2}_{\mathbb R} \mathrm{Mat}(2,\mathbb R) & p-q=0,2\mod 8 \\
\mathrm{Mat}(2^{(p+q-1)/2},\mathbb R)\oplus
\mathrm{Mat}(2^{(p+q-1)/2},\mathbb R)  & p-q=1\mod 8  \\
\mathrm{Mat}(2^{(p+q-1)/2},\mathbb C) & p-q=3,7\mod 8 \\
\mathrm{Mat}(2^{(p+q-2)/2},\mathbb H) & p-q=4,6\mod 8 \\
\mathrm{Mat}(2^{(p+q-3)/2},\mathbb H)\oplus
\mathrm{Mat}(2^{(p+q-3)/2},\mathbb H)  & p-q=5\mod 8
\end{array}\right. \label{k5}
\eeq
\end{theorem}

Since matrix algebras $\mathrm{Mat}(r,\cK)$, $\cK=\mathbb
R,\mathbb C, \mathbb H$, are simple, a glance at Table \ref{k5}
shows that real Clifford algebras $\cC(V,\eta)$ modelled over even
dimensional vector spaces $V$ (i.e., $p-q$ is even) are simple.

By a representation of a real Clifford algebra $\cC(V,\eta)$ is
meant its ring homomorphism $\rho$ to a real ring of linear
endomorphisms of a finite-dimensional real vector space $\Xi$,
whose dimension is called the dimension of a representation. A
representation is said to be exact if $\rho$ is an isomorphism. A
representation is called irreducible if there is no proper
subspace of $\Xi$ which is a carrier space of a representation of
$\cC(V,\eta)$.

Two representations $\rho$ and $\rho'$ of a Clifford algebra
$\cC(V,\eta)$ in vector spaces $\Xi$ and $\Xi'$ are said to be
equivalent if there is an isomorphism $\xi: \Xi\to \Xi'$ of these
vector spaces such that $\rho'=\xi\circ\rho\circ\xi^{-1}$ is a
real ring isomorphism of $\rho(\cC(V,\eta))$ and
$\rho'(\cC(V,\eta))$. The following is a corollary of Theorem
\ref{k33} \cite{law}.

\begin{theorem} \label{a10} \mar{a10}
If $n=\mathrm{dim}\,V$ is even, an exact irreducible
representation of a real ring $\cC(m,n-m)$ is unique up to an
equivalence. If $n$ is odd there exist two inequivalent exact
irreducible representations of a real Clifford algebra
$\cC(m,n-m)$.
\end{theorem}

Now, let us consider the complexification
\mar{63}\beq
\mathbb C\cC(m,n-m)=\mathbb C\op\otimes_{\mathbb R}\cC(m,n-m)
\label{63}
\eeq
of a real ring $\cC(m,n-m)$. It is readily observed that all
complexifications $\mathbb C\cC(m,n-m)$, $m=0,\ldots, n$, are
isomorphic:
\mar{66}\beq
\mathbb C\cC(m,n-m)= \mathbb C\cC(m',n-m'), \label{66}
\eeq
both as real and complex rings. Though the isomorphisms (\ref{66})
are not unique, one can speak about an abstract complex ring
$\mathbb C\cC(n)$ (\ref{66}) so that, given a real Clifford
algebra $\cC(m,n-m)$ and its complexification $\mathbb
C\cC(m,n-m)$ (\ref{63}), there exists the complex ring isomorphism
of $\mathbb C\cC(m,n-m)$ to $\mathbb C\cC(n)$. We call $\mathbb
C\cC(n)$ (\ref{66}) the \textbf{complex Clifford algebra}, and
define it as a complex ring
\mar{a25}\beq
\mathbb C\cC(n)=\mathbb C\op\otimes_{\mathbb R}\cC(n,0),
\label{a25}
\eeq
generated by $n$ elements $(e^i)$ such that
\mar{210}\beq
e^ie^j+e^je^i==2\kappa(e^i,e^j)e=2\delta^{ij}e. \label{210}
\eeq

Let us call $\{e^i\}$ (\ref{210}) the Euclidean basis for a
complex Clifford algebra $\mathbb C\cC(n)$. A complex vector space
$\cV$, spanned by an Euclidean basis $\{e^i\}$ and provided with
the bilinear form $\kappa$ (\ref{210}), is termed the Euclidean
generating space of a complex Clifford algebra $\mathbb C\cC(n)$.
With this basis, the complex ring $\mathbb C\cC(n)$ (\ref{a25})
possesses a canonical real subring
\mar{sp200}\beq
\cC(m,n-m)\to \mathbb C\cC(n) \label{sp200}
\eeq
with a basis $\{e^1,\ldots, e^m, ie^{m+1}, \ldots, ie^n\}$.

Theorem \ref{k33}  provides the following classification of
complex Clifford algebras $\mathbb C\cC(n)$ (\ref{a25})
\cite{law,cliff15}.

\begin{theorem} \label{k8} \mar{k8} Complex Clifford algebras are
isomorphic to the following matrix ones
\mar{k9}\beq
\mathbb C\cC(n) = \left\{
\begin{array}{ll}
\mathrm{Mat}(2^{n/2},\mathbb C)=\op\ot^{n/2}_{\mathbb C}
\mathrm{Mat}(2,\mathbb C)
=\op\ot^{n/2}_{\mathbb C} \mathbb C\cC(2) & n=0\mod 2\\
\mathrm{Mat}(2^{(n-1)/2},\mathbb C)\oplus
\mathrm{Mat}(2^{(n-1)/2},\mathbb C)  & n=1\mod 2
\end{array}\right.\label{k9}
\eeq
\end{theorem}

\begin{corollary} \label{k12} \mar{k12}
Since matrix algebras $\mathrm{Mat}(n,\mathbb C)$ are simple and
central (i.e., their center is proportional to the unit matrix),
complex Clifford algebras $\mathbb C\cC(n)$ of even $n$ are
central simple algebras.
\end{corollary}

By a representation of a complex Clifford algebra $\mathbb
C\cC(n)$ is meant its morphism $\rho$ to a complex algebra of
linear endomorphisms of a finite-dimensional complex vector space.
The following is a corollary of Theorem \ref{k8} \cite{law}.

\begin{theorem} \label{a11} \mar{a11}
If $n$ is even, an exact irreducible representation of a complex
Clifford algebra $\mathbb C\cC(n)$ is unique up to an equivalence.
If $n$ is odd there exist two inequivalent exact irreducible
representations of a complex Clifford algebra $\mathbb C\cC(n)$.
\end{theorem}

In view of Corollary \ref{k12} and Theorem \ref{a11}, we hereafter
focus our consideration on real and complex Clifford algebras
modelled over even vector spaces.

\subsection{Automorphisms of Clifford algebras}

We consider both generic ring automorphisms of a Clifford algebra
and its automorphisms which preserve a specified generating space
\cite{cliff15}.

Let $\cC(V,\eta)$ be a real Clifford algebra modelled over an
even-dimensional pseudo-Euclidean space $(V,\eta)$. By
Aut$[\cC(V,\eta)]$ is denoted the group of automorphisms of a real
ring $\cC(V,\eta)$. A key point is the following.

\begin{theorem} \label{k15} \mar{k15}
Any automorphism of a real ring $\cC(V,\eta)$ is inner.
\end{theorem}

Indeed, Theorem \ref{k33} states that any real Clifford algebra
$\cC(p,q)$, $p-q=0\mod 2$ as a ring is isomorphic to some matrix
algebra $\mathrm{Mat}(m,\cK)$, $\cK=\mathbb R,\mathbb C,\mathbb
H$. Such an algebra is simple. Algebras $\mathrm{Mat}(m,\cK)$,
$\cK=\mathbb R,\mathbb H$, are central simple real algebras with
the center $\cZ=\mathbb R$. Algebras $\mathrm{Mat}(m,\mathbb C)$
are central simple complex algebras with the center $\cZ=\mathbb
C$. In accordance with the well-known Skolem--Noether theorem
automorphisms of these algebras are inner.

Invertible elements of a real Clifford algebra
$\cC(V,\eta)=\mathrm{Mat}(m,\cK)$ constitute a general linear
matrix group $\cG\cC(V,\eta)=Gl(m,\cK)$. In particular, this group
contains all elements $v\in V\subset \cC(V,\eta)$ such that
$\eta(v,v)\neq 0$. Acting in $\cC(V,\eta)$ by left and right
multiplications, the group $\cG\cC(V,\eta)$ also acts in a real
Clifford algebra by the adjoint representation
\mar{a5}\beq
 \wh g: q\to gqg^{-1}, \qquad g\in \cG\cC(V,\eta),\qquad q\in
\cC(V,\eta). \label{a5}
\eeq
By virtue of Theorem \ref{k15}, this representation provides an
epimorphism
\mar{83}\beq
\zeta:\cG\cC(V,\eta)=Gl(m,\cK) \to
Gl(m,\cK)/\cZ=\mathrm{Aut}[\cC(V,\eta)]. \label{83}
\eeq

Any ring automorphism $g$ of $\cC(V,\eta)$ sends a generating
pseudo-Euclidean space $(V,\eta)$ of $\cC(V,\eta)$ onto an
isometrically isomorphic pseudo-Euclidean space $(V',\eta')$ such
that
\be
2\eta'(g(v),g(v'))e=g(v)g(v')+g(v')g(v)= 2\eta(v,v')e, \qquad
v,v'\in V.
\ee
It also is a generating space of a ring $\cC(V,\eta)$. Conversely,
let $(V,\eta)$ and $(V',\eta')$ be two different pseudo-Euclidean
generating spaces of the same signature of a ring $\cC(V,\eta)$.
In accordance with Lemma \ref{ss4}, their isometric isomorphism
$(V,\eta) \to (V',\eta')$ gives rise to an automorphism of a ring
$\cC(V,\eta)$ which also is an isomorphism of Clifford algebras
$\cC(V,\eta)\to \cC(V',\eta')$.

In particular, any (isometric) automorphism
\be
 g:V\ni v\to
g(v)\in V, \qquad \eta(g(v),g(v'))=\eta(v,v'), \qquad g\in
O(V,\eta),
\ee
of a pseudo-Euclidean generating space $(V,\eta)$ is prolonged to
an automorphism of a ring $\cC(V,\eta)$ which also is an
automorphism of a real Clifford algebra $\cC(V,\eta)$. Then we
have a monomorphism
\mar{82}\beq
O(V,\eta)\to \mathrm{Aut}[\cC(V,\eta)\,] \label{82}
\eeq
of a group $O(V,\eta)$ of automorphisms of a pseudo-Euclidean
space $(V,\eta)$ to a group of ring automorphisms of
$\cC(V,\eta)$. Herewith, an automorphism $g\in O(V,\eta)$ of a
ring $\cC(V,\eta)$ is the identity one iff its restriction to $V$
is an identity map of $V$. Consequently, the following is true.

\begin{theorem} \label{a3} \mar{a3}
A subgroup $O(V,\eta)\subset \mathrm{Aut}[\cC(V,\eta)\,]$
(\ref{82}) exhausts all automorphisms of a ring $\cC(V,\eta)$
which are automorphisms of a Clifford algebra $\cC(V,\eta)$.
\end{theorem}

Let us consider a subgroup $\mathrm{Cliff}(V,\eta)\subset
\cG\cC(V,\eta)$ generated by all invertible elements of $V\subset
\cC(V,\eta)$. It is called the Clifford group. One can show that
the homomorphism $\zeta$ (\ref{83}) of a Clifford group
$\mathrm{Cliff}(V,\eta)$ to $\mathrm{Aut}[\cC(V,\eta)]$ is its
epimorphism
\mar{103}\beq
\zeta: \cG\cC(V,\eta) \supset \mathrm{Cliff}(V,\eta)\to O(V,\eta)
\subset \mathrm{Aut}[\cC(V,\eta)] \label{103}
\eeq
onto $O(V,\eta)$. Due to the factorization (\ref{103}), any ring
automorphism $\wh v$, $v\in \mathrm{Cliff}(V,\eta)$, of
$\cC(V,\eta)$ also is an automorphism of a real Clifford algebra
$\cC(V,\eta)$.

The epimorphism (\ref{103}) yields an action of a Clifford group
$\mathrm{Cliff}(V,\eta)$ in a pseudo-Euclidean space $(V,\eta)$ by
the adjoint representation (\ref{a5}). However, this action is not
effective. Therefore, one consider subgroups Pin$(V,\eta)$ and
Spin$(V,\eta)$ of $\mathrm{Cliff}(V,\eta)$. The first one is
generated by elements $v\in V$ such that $\eta(v,v)=\pm 1$. A
group Spin$(V,\eta)$ is defined as an intersection
\mar{104a}\beq
\mathrm{Spin}(V,\eta)=\mathrm{Pin}(V,\eta)\cap \cC^0(V,\eta)
\label{104a}
\eeq
of a group Pin$(V,\eta)$ and the even subring $\cC^0(V,\eta)$ of a
real Clifford algebra $\cC(V,\eta)$. In particular, generating
elements $v\in V$ of Pin$(V,\eta)$ do not belong to its subgroup
Spin$(V,\eta)$. The epimorphism (\ref{103}) restricted to the Pin
and Spin groups leads to short exact sequences of groups
\mar{106,4}\ben
&& e\to \mathbb Z_2\longrightarrow
\mathrm{Pin}(V,\eta)\op\longrightarrow^\zeta O(V,\eta)\to e.
\label{106} \\
&& e\to \mathbb Z_2\longrightarrow
\mathrm{Spin}(V,\eta)\op\longrightarrow^\zeta SO(V,\eta)\to e,
\label{104}
\een
where $\mathbb Z_2\to (e,-e)\subset \mathrm{Spin}(V,\eta)$.

\begin{remark}
It should be emphasized that an epimorphism $\zeta$ in (\ref{106})
and (\ref{104}) is not a trivial bundle unless $\eta$ is of
signature $(1,1)$. It is a universal coverings over each component
of $O(V,\eta)$.
\end{remark}

Let $\mathbb C\cC(n)$ be the complex Clifford algebra (\ref{a25})
of even $n$.

\begin{theorem} \label{k21} \mar{k21}
All automorphisms of a complex Clifford algebra $\mathbb C\cC(n)$
are inner.
\end{theorem}

Indeed, by virtue of Theorem \ref{k8}, there is the ring
isomorphism (\ref{k9}):
\mar{k26}\beq
\mathbb C\cC(n)=\mathrm{Mat}(2^{n/2}, \mathbb C). \label{k26}
\eeq
In accordance with Corollary \ref{k12}, this algebra is a central
simple complex algebra with the center $\cZ=\mathbb C$.  In
accordance with the above-mentioned Skolem--Noether theorem
automorphisms of these algebras are inner. Invertible elements of
the Clifford algebra (\ref{k26}) constitute a general linear group
\mar{gg25}\beq
\cG\mathbb C\cC(n)= GL(2^{n/2},\mathbb C). \label{gg25}
\eeq
Acting in $\mathbb C\cC(n)$ by left and right multiplications,
this group also acts in a Clifford algebra by the adjoint
representation, and we obtain its epimorphism
\mar{k25}\ben
&& GL(2^{n/2},\mathbb C) \to \mathrm{Aut}[\cC(n)]=PGL(2^{n/2},\mathbb
C)= \label{k25}\\
&& \qquad GL(2^{n/2}, \mathbb
C)/\mathbb C=  SL(2^{n/2}, \mathbb C)/\mathbb Z_{2^{n/2}}
\nonumber
\een
onto a projective linear group $PGL(2^{n/2},\mathbb C)$.

Any automorphism $g$ of a complex Clifford algebra $\mathbb
C\cC(n)$ sends its Euclidean generating space $(\cV,\kappa)$ onto
some generating space
\be
(\cV',\kappa'), \qquad \kappa'(g(v),g(v'))=\kappa(v,v'), \qquad
v,v'\in \cV,
\ee
which is the Euclidean one with respect to the basis $\{g(e^i\}$.
Conversely, any automorphism of an Euclidean generating space
$(\cV,\kappa)$ is prolonged to an automorphism of a ring $\mathbb
C\cC(n)$. Then we have a monomorphism
\mar{ss25}\beq
O(n,\mathbb C)\to \mathrm{Aut}[\mathbb C\cC(n)] \label{ss25}
\eeq
of a group $O(n,\mathbb C)$ of automorphisms of an Euclidean
generating space $(\cV,\kappa)$ to a group of ring automorphisms
of $\mathbb C\cC(n)$. Herewith, an automorphism $g\in O(n,\mathbb
C)$ of a complex ring $\mathbb C\cC(n)$ is the identity one iff
its restriction to $\cV$ is an identity map of $\cV$.
Consequently, all ring automorphisms of a complex Clifford algebra
$\mathbb C\cC(n)$ preserving its Euclidean generating space form a
group $O(n,\mathbb C)$.

Given a complex Clifford algebra $\mathbb C\cC(n)$, let
$\cC(m,n-m)$ be a real Clifford algebra. Due to the canonical ring
monomorphism $\cC(m,n-m)\to \mathbb C\cC(n)$ (\ref{sp200}), there
is the canonical group monomorphism
\mar{sp201}\beq
\cG\cC(m,n-m)\to \cG\mathbb C\cC(n)=GL(2^{n/2},\mathbb C).
\label{sp201}
\eeq
Since all ring automorphisms of a real  Clifford algebra are inner
(Theorem \ref{k15}), they are extended to inner automorphisms of a
complex Clifford algebra $\mathbb C\cC(n)$.

\subsection{Spinor spaces}

As was mentioned above, we define spinor spaces in terms of
Clifford algebras \cite{book09,cliff15}.

A \textbf{real spinor space} $\Psi(m,n-m)$ is defined as a carrier
space of an irreducible representation of a real Clifford algebra
$\cC(m,n-m)$. It also carries out a representation of the
corresponding group Spin$(m,n-m)\subset \cC(m,n-m)$ \cite{law}.

If $n$ is even, such a real spinor space is unique up to an
equivalence in accordance with Theorem \ref{a10}. However, spinor
spaces $\Psi(m,n-m)$ and $\Psi(m',n-m')$ need not be isomorphic
vector spaces for $m'\neq m$.

A complex spinor space $\Psi(n)$ is defined as a carrier space of
an irreducible representation of a complex Clifford algebra
$\mathbb C\cC(n)$.

Since $n$ is even, a representation $\Psi(n)$ is unique up to an
equivalence in accordance with Theorem \ref{a11}. Therefore, it is
sufficient to describe a complex spinor space $\Psi(n)$ as a
subspace of a complex Clifford algebra $\mathbb C\cC(n)$ which
acts on $\Psi(n)$ by left multiplications.

Given a complex Clifford algebra $\mathbb C\cC(n)$, let us
consider its non-zero minimal left ideal which $\cC(n)$ acts on by
left multiplications. It is a finite-dimensional complex vector
space. Therefore, an action of a complex Clifford algebra $\mathbb
C\cC(n)$ in a minimal left ideal by left multiplications defines a
linear representation of $\mathbb C\cC(n)$. It obviously is
irreducible. In this case, a minimal left ideal of $\mathbb
C\cC(n)$ is a complex spinor space $\Psi(n)$. Thus, we come to an
equivalent definition of a \textbf{complex spinor space} as a
minimal left ideal of a complex Clifford algebra $\mathbb C\cC(n)$
which carry out its irreducible representation (\ref{sp110})
\cite{cliff15}.

By virtue of Theorem \ref{k8}, there is a ring isomorphism
$\mathbb C\cC(n)=\mathrm{Mat}(2^{n/2}, \mathbb C)$ (\ref{k26}).
Consequently, a spinor representation of a complex Clifford
algebra $\mathbb C\cC(n)$ is equivalent to the canonical
representation of $\mathrm{Mat}(2^{n/2}, \mathbb C)$ by matrices
in a complex vector space $\mathbb C^{2^{n/2}}$, i.e.,
$\Psi(n)=\mathbb C^{2^{n/2}}$. A spinor space $\Psi(n)\subset
\mathbb C\cC(n)$ also carries out the left-regular irreducible
representation of the group $\cG\mathbb C\cC(n)=GL(2^{n/2},\mathbb
C)$ (\ref{gg25}) which is equivalent to the natural matrix
representation of $GL(2^{n/2},\mathbb C)$ in $\mathbb
C^{2^{n/2}}$.

Owing to the monomorphism $\cC(m,n-m)\to \mathbb C\cC(n)$
(\ref{sp200}), a spinor space $\Psi(n)$ also carries out a
representation of real Clifford algebras $\cC(m,n-m)$, their Pin
and Spin groups, though these representation need not be
reducible.

\begin{remark} \label{gg15} \mar{gg15}
Certainly, an automorphism of a Clifford algebra $\mathbb C\cC(n)$
sends a spinor space onto a spinor space, but not the same one. An
action of a group $PGL(2^{n/2},\mathbb C)$ of automorphisms of
$\mathbb C\cC(n)$ in a set $S\Psi(n)$ of spinor spaces is
transitive.
\end{remark}

\subsection{Clifford algebra bundles and spinor bundles}

Treating spinor spaces as subspaces of Clifford algebras, we can
describe spinor bundles as subbundles of a fibre bundle in complex
Clifford algebras \cite{cliff15}.

One usually consider fibre bundles in Clifford algebras whose
structure group is a group of automorphisms of these algebras
\cite{book09,law}. A problem is that, as was mentioned above, this
group fails to preserve spinor subspaces of a complex Clifford
algebra (Remark \ref{gg15}) and, thus, it can not be a structure
group of spinor bundles. Therefore, we define fibre bundles in
Clifford algebras whose structure group is a group of invertible
elements of a complex Clifford algebra which acts on this algebra
by left multiplications. Certainly, it preserves minimal left
ideals of this algebra and, consequently, it is a structure group
of spinor bundles.

Let $\mathbb C\cC(n)$ be a complex Clifford algebra modelled over
an even dimensional complex space $\mathbb C^n$. It is isomorphic
to a ring $\mathrm{Mat}(2^{n/2}, \mathbb C)$ of complex
$(2^{n/2}\times 2^{n/2})$-matrices (Theorem \ref{k8}). Its
invertible elements constitute the general linear group
$\cG\mathbb C\cC(n)=GL(2^{n/2},\mathbb C)$ (\ref{gg25}) whose
adjoint representation in $\mathbb C\cC(n)$ yields the projective
linear group $PGL(2^{n/2}, \mathbb C)$ (\ref{k25}) of
automorphisms of $\mathbb C\cC(n)$ (Theorem \ref{k21}).

Given a smooth manifold $X$, let us consider a principal bundle
$P\to X$ with a structure group $GL(2^{n/2}, \mathbb C)$. A fibre
bundle in complex Clifford algebras $\mathbb C\cC(n)$ is defined
to be the $P$-associated bundle:
\mar{sp100}\beq
\lC= (P\times \mathrm{Mat}(2^{n/2}, \mathbb C))/GL(2^{n/2},
\mathbb C)\to X \label{sp100}
\eeq
with a typical fibre $\mathbb C\cC(n)= \mathrm{Mat}(2^{n/2},
\mathbb C)$ which carries out the left-regular representation of a
group $GL(2^{n/2}, \mathbb C)$.

Owing to the canonical inclusion $GL(2^{n/2}, \mathbb C)\to
\mathrm{Mat}(2^{n/2}, \mathbb C)$, a principal $GL(2^{n/2},
\mathbb C)$-bundle $P$ is a subbundle $P\subset \lC$ of the
Clifford algebra bundle $\lC$ (\ref{sp100}). Herewith, the
canonical right action of a structure group $GL(2^{n/2}, \mathbb
C)$ on a principal bundle $P$ is extended to the fibrewise action
of $GL(2^{n/2}, \mathbb C)$ on the Clifford algebra bundle $\lC$
(\ref{sp100}) by right multiplications. This action is globally
defined because it is commutative with transition functions of
$\lC$ acting on its typical fibre $\mathrm{Mat}(2^{n/2}, \mathbb
C)$ on the left.

\begin{remark} \label{sp103} \mar{sp103}
As was mentioned above, one usually considers a fibre bundle in
Clifford algebras $\mathbb C\cC(n)=\mathrm{Mat}(2^{n/2})$
(\ref{k26}) whose structure group is the group $PGL(2^{n/2},
\mathbb C)$ (\ref{k25}) of automorphisms of $\mathbb C\cC(n)$.
This also is a $P$-associated bundle
\mar{sp104}\beq
\cA\lC=(P\times \mathbb C\cC(n))/GL(2^{n/2}, \mathbb C)\to X
\label{sp104}
\eeq
where $GL(2^{n/2}, \mathbb C)$ acts on $\mathbb C\cC(n)$ by the
adjoint representation.
\end{remark}

Let $\Psi(n)$ be a spinor space of a complex Clifford algebra
$\mathbb C\cC(n)$. Being a minimal left ideal of $\mathbb
C\cC(n)$, it is a subspace $\Psi(n)$ of $\mathbb C\cC(n)$ which
inherits the left-regular representation of a group $GL(2^{n/2},
\mathbb C)$ in $\mathbb C\cC(n)$. Given a principal $GL(2^{n/2},
\mathbb C)$-bundle $P$, a \textbf{spinor bundle} then is defined
as a $P$-associated bundle
\mar{sp107}\beq
S=(P\times \Psi(n))/GL(2^{n/2}, \mathbb C)\to X  \label{sp107}
\eeq
with a typical fibre $\Psi(n)=\mathbb C^{2^{n/2}}$ and a structure
group $GL(2^{n/2}, \mathbb C)$ which acts on $\Psi(n)$ by left
multiplications.

Obviously, the spinor bundle $S$ (\ref{sp107}) is a subbundle of
the Clifford algebra bundle $\lC$ (\ref{sp100}). However, $S$
(\ref{sp107}) need not  be a subbundle of the fibre bundle
$\cA\lC$ (\ref{sp104}) in Clifford algebras because a spinor space
$\Psi(n)$ is not stable under automorphisms of a complex Clifford
algebra $\mathbb C\cC(n)$.

At the same time, given the spinor representation  (\ref{sp110})
of a complex Clifford algebra, there is a fibrewise representation
morphism
\mar{sp111}\ben
&& \g: \cA\lC\op\times_X S\ar_X S,  \label{sp111}\\
&& \g: (P\times (\mathbb C\cC(n)\times \Psi(n)))/\cG\mathbb C\cC(n)
\to
 (P\times \g(\mathbb C\cC(n)\times \Psi(n)))/\cG\mathbb C\cC(n), \nonumber
\een
of the $P$-associated fibre bundles $\cA\lC$ (\ref{sp104}) and $S$
(\ref{sp107}) with a structure group $\cG\mathbb C\cC(n)$.

It should be emphasized that, though there is the ring
monomorphism $\cC(m,n-m)\to \mathbb C\cC(n)$ (\ref{sp200}), the
Clifford algebra bundle $\lC$ (\ref{sp100}) need not contains a
subbundle in real Clifford algebras $\cC(m,n-m)$, unless a
structure group $GL(2^{n/2}, \mathbb C)$ of $\lC$ is reducible to
a subgroup $\cG\cC(m,n-m)$. This problem can be solved as follows.

Let $X$ be a smooth real manifold of even dimension $n$. Let
$T^*X$ be the cotangent bundle over $X$ and $LX$ the associated
principal frame bundle. Let us assume that their structure group
is $GL(n, \mathbb R)$ is reducible to a pseudo-ortohogonal
subgroup $O(m,n-m)$. In particular, a structure group $GL(n,
\mathbb R)$ always is reducible to a maximal compact subgroup
$O(n,\mathbb R)$ (Theorem \ref{gg21}). There is the exact sequence
of groups (\ref{106}):
\mar{sp121}\beq
 e\to \mathbb Z_2\longrightarrow
\mathrm{Pin}(m,n-m)\op\longrightarrow^\zeta O(m,n-m)\to e.
\label{sp121}
\eeq
A problem is that this exact sequence need not be split, i.e.,
there is no monomorphism $\kappa: O(m,n-m)\to \mathrm{Pin}(m,n-m)$
so that $\zeta\circ\kappa=\id$, in general.

In this case, we say that a principal $\mathrm{Pin}(m,n-m)$-bundle
$\wt Y\to X$ is an extension of a principal $O(m,n-m)$-bundle
$Y\to X$ if there is an epimorphism of principal bundles
\mar{sp223}\beq
\wt Y\ar_X Y. \label{sp223}
\eeq
Such an extension need not exist.

\begin{remark} \label{gg23} \mar{gg23}
The topological obstruction to that a principal $O(m,n-m)$-bundle
$Y\to X$ lifts to a principal $\mathrm{Pin}(m,n-m)$-bundle $\wt
Y\to X$ is given by the \v Cech cohomology group $H^2(X;\mathbb
Z_2)$ of $X$ \cite{book09,greub,law}. Namely, a principal bundle
$Y$ defines an element of $H^2(X;\mathbb Z_2)$ which must be zero
so that $Y\to X$ can give rise to $\wt Y\to X$. Inequivalent lifts
of $Y\to X$ to principal $\mathrm{Pin}(m,n-m)$-bundles are
classified by elements of the \v Cech cohomology group
$H^1(X;\mathbb Z_2)$.
\end{remark}

 Let $L^hX$ be a reduced principal
$O(m,n-m)$-subbundle of a frame bundle $LX$. In this case, the
above mentioned topological obstruction to that this bundle $L^hX$
is extended to a principal $\mathrm{Pin}(m,n-m)$-bundle $\wt L^hX$
(Remark \ref{gg23}) is the second Stiefel--Whitney class
$w_2(X)\in H^2(X;\mathbb Z_2)$ of $X$ \cite{law}. Let us assume
that a manifold $X$ is orientable, i.e., the \v Cech cohomology
group $H^1(X;\mathbb Z_2)$ is trivial,  and that the second
Stiefel--Whitney class $w_2(X)\in H^2(X;\mathbb Z_2)$ of $X$ also
is trivial. Let $\wt L^hX$ be the desired
$\mathrm{Pin}(m,n-m)$-lift (\ref{sp223}) of a principal
$O(m,n-m)$-bundle $L^hX$. Owing to the canonical monomorphism
(\ref{sp200}) of Clifford algebras, there is the group
monomorphism $\mathrm{Pin}(m,n-m)\to \cG\mathbb C\cC(n)$
(\ref{sp201}). Due to this monomorphism, there exists a principal
$\cG\mathbb C\cC(n)$-bundle $P^h$ whose reduced
$\mathrm{Pin}(m,n-m)$-subbundle is $\wt L^hX$, and whose structure
group $\cG\mathbb C\cC(n)=GL(2^{n/2}, \mathbb C)$ (\ref{gg25})
thus is reducible to $\mathrm{Pin}(m,n-m)$. Let
\mar{sp240}\beq
\lC^h= (P^h\times \mathrm{Mat}(2^{n/2}, \mathbb C))/GL(2^{n/2},
\mathbb C)\to X \label{sp240}
\eeq
be the $P^h$-associated bundle (\ref{sp100}) in complex Clifford
algebras $\mathbb C\cC(n)$. Then it contains a subbundle
\mar{sp241}\beq
\lC^h(m,n-m)=(\wt L^hX\times \cC(m,n-m))/\mathrm{Pin}(m,n-m)\to
X\label{sp241}
\eeq
in real Clifford algebras $\cC(m,n-m)$. The Clifford algebra
bundle $\lC^h$ (\ref{sp240}) also contains spinor subbundles
(\ref{sp107}):
\mar{gg30}\beq
S^h=(P^h\times \Psi(4))/GL(2^{n/2},\mathbb C)\to X. \label{gg30}
\eeq

Let us consider a $P^h$-associated fibre bundle $\cA\lC^h$
(\ref{sp104}) in complex Clifford algebras $\mathbb C\cC(n)$ whose
structure group acts on $\mathbb C\cC(n)$ by the adjoint
representation and, thus, it is the group $\mathrm{Aut}[\cC(n)]$
(\ref{k25}) of its automorphisms. Since a structure group of $P^h$
is reducible to $\mathrm{Pin}(m,n-m)$, a fibre bundle $\cA\lC^h$
contains a $\wt L^hX$-associated subbundle
\mar{gg31}\beq
\cA\lC^h(m,n-m)= (\wt L^hX\times
\cC(m,n-m))/\mathrm{Pin}(m,n-m)\to X \label{gg31}
\eeq
in real Clifford algebras $\cC(m,n-m)$ where the group
$\mathrm{Pin}(m,n-m)$ acts on $\cC(m,n-m)$ by the adjoint
representation. Then there is the fibrewise representation
(\ref{sp111}):
\mar{gg32}\beq
\g:\cA\lC^h(m,n-m)\op\times_X S^h\ar_X S^h. \label{gg32}
\eeq

Due to the epimorphism $\zeta$ (\ref{sp121}), the Clifford algebra
bundle $\cA\lC^h(m,n-m)$ contains a subbundle $M^hX$ in
pseudo-Euclidean generating spaces of fibres of
$\cA\lC^h(m,n-m)$with a structure group $O(m,n-m)$. It is
associated to an original reduced principal subbundle $L^hX$ of a
frame bundle  $LX$ and, thus, is isomorphic to the cotangent
bundle $T^*X$ of  $X$. Accordingly, the fibrewise representation
$\g$ (\ref{gg32}) leads to a  fibrewise Clifford algebra
representation
\mar{gg33}\beq
\g:M^hX\op\times_X S^h\ar_X S^h \label{gg33}
\eeq
of elements of the cotangent bundle $TX=M^hX$.

Of course, with a different reduced principal $O(m,n-m)$-subbundle
$L^{h'}X$ of $LX$, we come to a different Clifford algebra bundle
$\lC^{h'}$ (\ref{sp240}). By virtue of Theorem \ref{red}, there is
one-to-one correspondence between the reduced principal
$O(m,n-m)$-subbundle $L^hX$ of $LX$ and the global sections $h$ of
the quotient bundle
\mar{sp226}\beq
\Si(m,n-m)=LX/O(m,n-m)\to X, \label{sp226}
\eeq
which are pseudo-Riemannian metrics of signature $(m,n-m)$ on $X$.

\begin{remark} \label{gg20} \mar{gg20}
A key point is that, given different global sections $h$ and $h'$
of the quotient bundle $\Si(m,n-m)$ (\ref{sp226}), neither complex
Clifford algebra bundles $\lC^h$ and $\lC^{h'}$ (\ref{sp240}) nor
real Clifford algebra bundles $\cA\lC^h(m,n-m)$ and
$\cA\lC^{h'}(m,n-m)$ are not isomorphic. These fibre bundles are
associated to principal $\mathrm{Pin}(m,n-m)$-bundles $\wt L^hX$
and $\wt L^{h'}X$ which are the two-fold covers (\ref{sp223}) of
the reduced principal $O(m,n-m)$-subbundles $L^hX$ and $L^{h'}X$
of a frame bundle $LX$, respectively. These subbundles need not be
isomorphic, and then the principal bundles $\wt L^hX$ and $\wt
L^{h'}X$ are well. Moreover, let principal bundles $L^hX$ and
$L^{h'}X$ be isomorphic. For instance, this is the case of an
orthogonal group $O(n,\mathbb R)$. However, their covers $\wt
L^hX$ and $\wt L^{h'}X$ need not be isomorphic. An isomorphism of
$L^hX$ and $L^{h'}X$ yields an isomorphism of fibre bundles $M^hX$
and $M^{h'}X$ in generating pseudo-Euclidean spaces, but it is not
isometric, and, therefore, fails to provide an isomorphism of real
Clifford algebra bundles $\cA\lC^h(m,n-m)$ and
$\cA\lC^{h'}(m,n-m)$. Consequently, a Clifford algebra bundle must
be considered only in a pair with a certain pseudo-Riemannian
metric $h$.
\end{remark}

In order to describe a whole family of non-isomorphic Clifford
algebra bundles $\lC^h$, let us call into play a composite bundle
\mar{sp230}\beq
LX \ar_X \Si(m,n-m) \ar X \label{sp230}
\eeq
where
\mar{sp231}\beq
LX\ar_X \Si(m,n-m) \label{sp231}
\eeq
is a principal bundle with a structure group $O(m,n-m)$
\cite{cliff15}. Let us consider its principal
$\mathrm{Pin}(m,n-m)$-lift (\ref{sp223}):
\mar{sp232}\beq
\wt LX\ar_X \Si(m,n-m), \label{sp232}
\eeq
if this exists. It is a composite bundle
\mar{gg40}\beq
\wt LX \ar_X \Si(m,n-m) \ar X. \label{gg40}
\eeq
Then, given a global section $h$ of $\Si(m,n-m) \to X$
(\ref{sp226}), the pull-back $h^* LX$ of $LX\to \Si(m,n-m)$
(\ref{sp231}) is a reduced principal $O(m,n-m)$-subbundle $L^hX$
of the frame bundle $LX\to X$ (\ref{sp230}). Accordingly, the
pull-back $h^* \wt LX$ of $\wt LX\to \Si(m,n-m)$ (\ref{sp232}) is
a principal $\mathrm{Pin}(m,n-m)$-subbundle of the composite
bundle $\wt LX\to X$ (\ref{gg40}), and it is a
$\mathrm{Pin}(m,n-m)$-lift
\mar{gg41}\beq
h^*\wt LX=\wt L^hX \ar_X L^hX \label{gg41}
\eeq
of $L^hX=h^*LX$.

Owing to the group monomorphism $\mathrm{Pin}(m,n-m)\to \cG\mathbb
C\cC(n)$ (\ref{sp201}), there exists a principal $\cG\mathbb
C\cC(n)$-bundle
\mar{sp245}\beq
P_\Si\ar_X \Si(m,n-m), \label{sp245}
\eeq
whose reduced principal $\mathrm{Pin}(m,n-m)$-subbundle is the
fibre bundle (\ref{sp232}). Let
\mar{sp246}\beq
\lC_\Si\ar_X \Si(m,n-m) \label{sp246}
\eeq
be the $P_\Si$-associated bundle (\ref{sp100}) in complex Clifford
algebras $\mathbb C\cC(n)$. It contains a $\wt LX$-associated
subbundle
\mar{sp247}\beq
\lC_\Si(m,n-m)\ar_X \Si(m,n-m)\label{sp247}
\eeq
in real Clifford algebras $\cC(m,n-m)$. The Clifford algebra
bundle $\lC_\Si$ (\ref{sp246}) also has $P_\Si$-associated spinor
subbundles
\mar{sp248}\beq
S_\Si\ar_X \Si(m,n-m). \label{sp248}
\eeq

Given a global section $h$ of $\Si(m,n-m) \to X$ (\ref{sp226}),
the pull-back bundles $h^*\lC_\Si\to X$, $h^*\lC_\Si(m,n-m)\to X$
and $h^*S_\Si\to X$ are subbundles of the composite bundles
$\lC_\Si\to X$, $\lC_\Si(m,n-m)\to X$ and $S_\Si\to X$ and are the
bundles $\lC^h\to X$ (\ref{sp240}), $\lC^h(m,n-m)\to X$
(\ref{sp241}) and $S^h\to X$ (\ref{gg30}), respectively.

Similarly, we define an $\wt LX$-associated bundle
\mar{gg45}\beq
\cA\lC_\Si(m,n-m) \ar_X \Si(m,n-m) \label{gg45}
\eeq
in real Clifford algebras $\cC(m,n-m)$ where the group
$\mathrm{Pin}(m,n-m)$ acts on $\cC(m,n-m)$ by the adjoint
representation. Then there is a fibrewise Clifford algebra
representation
\mar{gg46}\beq
\g:\cA\lC_\Si(m,n-m)\op\times_\Si S_\Si\ar_\Si S_\Si. \label{gg46}
\eeq
Given a global section $h$ of $\Si(m,n-m) \to X$ (\ref{sp226}),
the pull-back bundle $h^*\cA\lC_\Si(m,n-m)\to X$ restarts the
Clifford algebra bundle $\cA\lC^h(m,n-m)$ (\ref{gg31}) and the
fibrewise representation $\g$ (\ref{gg32}).

Due to the epimorphism $\zeta$ (\ref{sp121}), the Clifford algebra
bundle $\cA\lC_\Si(m,n-m)$ (\ref{gg45}) contains a subbundle
$M_\Si$ in pseudo-Euclidean generating spaces of fibres of
$\cA\lC_\Si(m,n-m)$ with a structure group $O(m,n-m)$. This fibre
bundle is associated to the principal $O(m,n-m)$-bundle
(\ref{sp231}), and it inherits the fibrewise representation
(\ref{gg46}):
\mar{gg47}\beq
\g:M_\Si\op\times_\Si S_\Si\ar_\Si S_\Si. \label{gg47}
\eeq
Given a global section $h$ of $\Si(m,n-m) \to X$ (\ref{sp226}),
its pull-back $h^*M_\Si\to X$ coincides with a fibre bundle $M^hX$
in pseudo-Euclidean generating spaces, and it is isomorphic to the
cotangent bundle $T^*X$ of $X$. Accordingly, the fibrewise
representation $\g$ (\ref{gg47}) reproduces that (\ref{gg33}).

\section{Dirac spinor fields in gauge gravitation theory}

A Dirac spinor space is defined to be a spinor space $\Psi(1,3)$
of an irreducible representation of real Clifford algebra
$\cC(1,3)$.

There are ring isomorphisms of real Clifford algebras
\mar{11}\beq
\cC(1,3)= \cC(4,0)= \cC(0,4)=  \mathrm{Mat}(2,\mathbb H),
\label{11}
\eeq
which as rings fail to be isomorphic to real Clifford algebras
\mar{12}\beq
\cC(3,1)=\cC(2,2)= \mathrm{Mat}(4,\mathbb R). \label{12}
\eeq
Due to the isomorphism (\ref{12}), a real Clifford algebra
$\cC(3,1)$ possesses an irreducible four-dimensional
representation by real matrices
\mar{62}\beq
\begin{pmatrix} 0 & \mathbf{1} \\ \mathbf{1} & 0
\end{pmatrix}, \qquad
\begin{pmatrix} 0 & -\mathbf{1} \\ \mathbf{1} & 0
\end{pmatrix}, \qquad
 \begin{pmatrix} \sigma^1 & 0
\\ 0 & -\sigma^1 \end{pmatrix} , \qquad \begin{pmatrix} \sigma^3 & 0
\\ 0 & -\sigma^3 \end{pmatrix} , \label{62}
\eeq
where $\mathbf{1}$ is the unit $(2\times 2)$-matrix and
$\sigma^k$, $k=1,2,3$, are Pauli matrices
\mar{52}\beq
\sigma^1=\begin{pmatrix}0 & 1\\1 & 0 \end{pmatrix}, \qquad
\sigma^2=\begin{pmatrix}0 & -i\\i & 0\end{pmatrix}, \qquad
\sigma^3=\begin{pmatrix}1 & 0\\0 & -1\end{pmatrix}. \label{52}
\eeq
By virtue of Theorem \ref{a10}, the representation (\ref{62}) is
unique up to an equivalence. Its carrier space is $\Psi(3,1)$ of
real \textbf{Majorana spinors}.

In contrast with the representation (\ref{62}) of $\cC(3,1)$, a
representation of a real Clifford algebra $\cC(3,1)$, the matrix
representation $\cC(1,3)= \mathrm{Mat}(2,\mathbb H)$ (\ref{11}) by
Dirac's $\g$-matrices
\mar{41'}\beq
\gamma^0=\begin{pmatrix}0 & \mathbf{1} \\ \mathbf{1} & 0
\end{pmatrix}, \qquad \gamma^j= \begin{pmatrix}0 &
-\sigma^j\\\sigma^j & 0 \end{pmatrix} \label{41'}
\eeq
is not real. As was mentioned above, we therefore consider complex
spinors which form a carrier space $\Psi(4)$ of an irreducible
representation of a complex Clifford algebra $\mathbb C\cC(4)$.
This representation is unique up to an equivalence in accordance
with Theorem \ref{a11}.

Let $\{e^0,e^i\}$  be the Euclidean basis (\ref{210}) for a
complex Clifford algebra $\mathbb C\cC(4)$. With this basis, the
complex ring $\mathbb C\cC(4)$ possesses a canonical real subring
$\cC(1,3)$ (\ref{sp200}) with a basis $\{e^0,ie^i\}$. Then
$\Psi(4)$ admits a representation of a complex Clifford algebra
$\mathbb C\cC(4)$ by the matrices $e^0=\g^0$, $e^i=-i\g^i$ whose
restriction to a real Clifford algebra $\cC(1,3)$ restarts its
representation (\ref{41'}) and provides a representation of a
group Spin$(1,3)$.

A group Spin$(1,3)$ contains two connected components
Spin$^+(1,3)$ and Spin$^-(1,3)$. Being a connected component of
the unity, the first one is a group $SL(2,\mathbb C)$. We have the
exact sequence (\ref{104}):
\be
e\to \mathbb Z_2\longrightarrow
\mathrm{Spin}(1,3)\op\longrightarrow^\zeta SO(1,3)\to e.
\ee
It is restricted to the exact sequence
\mar{105}\beq
e\to \mathbb Z_2\longrightarrow
\mathrm{Spin}^+(1,3)\op\longrightarrow^\zeta \mathrm{L} \to e,
\label{105}
\eeq
where a proper Lorentz group L is a connected component of the
unit of $SO(1,3)$. Let us call
\mar{300}\beq
\mathrm{L_s}=\mathrm{Spin}^+(1,3)=SL(2, \mathbb C) \label{300}
\eeq
the \textbf{Lorentz spin group}.

Group spaces of $\mathrm{L_s}$ and L are topological spaces
$S^3\times \mathbb R^3$ and $\mathbb R P^3\times \mathbb R^3$,
respectively. Their Lie algebras coincide with each other. It can
be provided with a basis $\{I_{ab}=-I_{ba}\}$, $a,b=0,1,2,3$ whose
elements obey the commutation relations
\be
[I_{ab}, I_{cd}]=\eta_{ad}I_{bc} + \eta_{bc}I_{ad} -
\eta_{ac}I_{bd} - \eta_{bd}I_{ac},
\ee
where $\eta$ is the Minkowski metric. Its representation
(\ref{41'}) in $\Psi(4)$ reads
\mar{gg70}\beq
I_{ab}=\frac14[\g_a,\g_b]. \label{gg70}
\eeq

Let $X$ be a world manifold. Let us assume that the second
Stiefel--Whitney class $w_2(X)\in H^2(X;\mathbb Z_2)$ of $X$ is
trivial (Remark \ref{gg23}). We follow the procedure in Section 10
in order to describe a Dirac spinor structure on $X$
\cite{book09,sard98,sard11}.

For this purpose, let us assume that the structure group $GL_4$
(\ref{gg1}) of a linear frame bundle $LX$ is reducible to a proper
Lorentz group $L$. By virtue of Theorem \ref{red}, there is
one-to-one correspondence between the principal L-subbundles
$L^hX$ of a frame bundle $LX$ and the global sections $h$ of the
quotient bundle $\Si_{\mathrm T}\to X$ (\ref{5.15}) called the
tetrad fields. Let us consider the composite bundle (\ref{sp230}):
\mar{gg55}\beq
LX\ar_X \Si_\mathrm{T}\ar X, \label{gg55}
\eeq
where
\mar{gg56}\beq
LX\ar_X \Si_\mathrm{T} \label{gg56}
\eeq
is a principal bundle with a structure group L. Given a tetrad
field $h$, the pull-back $h^* LX$ of $LX\to \Si_\mathrm{T}$
(\ref{gg56}) is a reduced principal L-subbundle $L^hX$ of a frame
bundle $LX\to X$.

Let us note that a structure group $GL_4$ of a frame bundle $LX$
is not simply-connected. Its first homotopy group is
\be
\pi_1(GL_4)= \pi_1({\rm SO}(4)) =\mathbb Z_2
\ee
\cite{green}. Therefore, a group $GL_4$ also admits the universal
two-fold covering group $\wt{GL}_4$  such that the diagram
\mar{pop1}\beq
\begin{array}{ccc}
 \wt{GL}_4 & \longrightarrow &  GL_4 \\
 \put(0,-10){\vector(0,1){20}} &
& \put(0,-10){\vector(0,1){20}}  \\
\mathrm{L_s} & \ar^\zeta & \rL
\end{array} \label{pop1}
\eeq
is commutative \cite{hehl,law,swit}.

\begin{remark}
Though a group $\wt{GL}_4$ admits finite-dimensional
representations, its fundamental spinor representation is
infinite-dimensional \cite{hehl,nee}. Elements of this
representation are called \textbf{world spinors}.  Their field
model has been developed (see \cite{hehl} and references therein).
\end{remark}

Given a group $\wt{GL}_4$, there exists a unique principal
$\wt{GL}_4$-bundle $\wt LX\to X$ which is a two-fold cover
\mar{pop}\beq
\wt LX  \ar_X^z  LX \label{pop}
\eeq
of a frame bundle $LX$. Due to the commutative diagram
(\ref{pop1}), there is a commutative diagram of principal bundles
\be
\begin{array}{ccc}
 \wt LX & \ar &  LX \\
 \put(0,10){\vector(1,-1){20}} &
& \put(0,10){\vector(-1,-1){20}}  \\
 & \Si_\mathrm{T} &
\end{array}
\ee
where
\mar{gg61}\beq
\wt LX \ar_X \wt LX/\mathrm{L_s}=\Si_\mathrm{T} \label{gg61}
\eeq
is a principal $\mathrm{L_s}$-bundle. It is just the
$\mathrm{L_s}$-lift (\ref{sp232}) of the principal L-bundle
$LX\to\Si_\mathrm{T}$ (\ref{gg56}).

Let us consider the composite bundle (\ref{gg40}):
\mar{gg65}\beq
\wt LX \ar_X \Si_\mathrm{T}\ar X. \label{gg65}
\eeq
Given a tetrad field $h$, the pull-back $h^* \wt LX$ of $\wt LX\to
\Si_\mathrm{T}$ (\ref{gg61}) is a reduced principal
$\mathrm{L_s}$-subbundle $\wt L^hX$ of the composite bundle $\wt
LX\to X$ (\ref{gg65}). Due to the commutative diagram
(\ref{pop1}), there is a commutative diagram of principal bundles
\mar{gg79}\beq
\begin{array}{ccc}
 \wt{LX} & \ar^z & LX \\
 \put(0,-10){\vector(0,1){20}} &
& \put(0,-10){\vector(0,1){20}}  \\
\wt L^hX & \ar^{z_h} & L^hX
\end{array} \label{gg79}
\eeq

Owing to the group monomorphism (\ref{105}):
\be
\mathrm{L_s}\to \cG\mathbb C\cC(4)=GL(4,\mathbb C),
\ee
there exists a principal $GL(4,\mathbb C)$-bundle
\mar{sp245a}\beq
P_\Si\ar_X \Si_\mathrm{T}, \label{sp245a}
\eeq
whose reduced principal $\mathrm{L_s}$-subbundle is the fibre
bundle (\ref{gg61}). Let
\mar{sp246a}\beq
\lC_\Si\ar_X \Si_\mathrm{T} \label{sp246a}
\eeq
be the $P_\Si$-associated bundle (\ref{sp100}) in complex Clifford
algebras $\mathbb C\cC(4)$. It contains a
$\mathbb{L_s}$-associated subbundle
\mar{sp247a}\beq
\lC_\Si(1,3)\ar_X \Si_\mathrm{T} \label{sp247a}
\eeq
in real Clifford algebras $\cC(1,3)$. The Clifford algebra bundle
$\lC_\Si$ (\ref{sp246a}) also has $P_\Si$-associated spinor
subbundles
\mar{sp248a}\beq
S_\Si\ar_X \Si_\mathrm{T} \label{sp248a}
\eeq
with a typical fibre $\Psi(4)$.

Similarly, we define a $\wt LX$-associated bundle
\mar{gg80}\beq
\cA\lC_\Si(1,3) \ar_X \Si_\mathrm{T} \label{gg80}
\eeq
in real Clifford algebras $\cC(1,3)$ where a group $\mathrm{L_s}$
acts on $\cC(1,3)$ by the adjoint representation. Then there is a
fibrewise representation morphism
\mar{gg81}\beq
\g:\cA\lC_\Si(1,3)\op\times_{\Si_\mathrm{T}}
S_\Si\ar_{\Si_\mathrm{T}} S_\Si. \label{gg81}
\eeq

Due to the epimorphism $\zeta$ (\ref{105}), the Clifford algebra
bundle $\cA\lC_\Si(1,3)$ (\ref{gg80}) contains a subbundle $M_\Si$
x in Minkowski generating spaces $\mathbb R^4\subset \cC(1,3)$
with a structure group L. This fibre bundle
\mar{gg90}\beq
M_\Si=(\wt LX\times \mathbb R^4)/\mathrm{L_s}=(LX\times \mathbb
R^4)/\mathrm{L} \label{gg90}
\eeq
is associated to the principal L-bundle (\ref{gg56}), and it
inherits the fibrewise representation (\ref{gg81}):
\mar{gg82}\beq
\g:M_\Si\op\times_{\Si_\mathrm{T}} S_\Si\ar_{\Si_\mathrm{T}}
S_\Si. \label{gg82}
\eeq

Given a tetrad field $h$,
\mar{gg78}\beq
S^h=h^*S_\Si\to X \label{gg78}
\eeq
of the spinor bundle $S_\Si$ (\ref{sp248a}) is a subbundle of a
composite bundle
\mar{ggz}\beq
S=S_\Si\ar_X \Si_{\mathrm T}\to X \label{ggz}
\eeq
and, in view of the commutative diagram (\ref{gg79}), it is a $\wt
L^hX$-associated bundle with the structure Lorentz spin group
$\mathrm{L_s}$ (\ref{300}).

With a tetrad field $h$, let us consider the pull-back
\mar{gg88}\beq
\cA\lC^h(1,3) =h^*\cA\lC^(1,3)\to X \label{gg88}
\eeq
of the Clifford algebra bundle $\cA\lC_\Si(1,3)$ (\ref{gg81}). It
contains the pull-back bundle
\mar{gg93}\beq
M^hX=h^*M_\Si=(L^hX\times \mathbb R^4)/\mathrm{L} \label{gg93}
\eeq
of generating Minkowski spaces. It is isomorphic to the cotangent
bundle
\be
T^*X=(L^hX\times \mathbb R^4)/\mathrm{L}
\ee
of $X$ if it is endowed the Lorentz atlas $\Psi^h$ (\ref{lat}).
The fibre bundle $\cA\lC^h(1,3)$ (\ref{gg88}) inherits the
fibrewise representation (\ref{gg81}):
\mar{gg95}\beq
\g_h:\cA\lC^h(1,3)\op\times_X S^h\ar_X S^h, \label{gg95}
\eeq
and $M^hX$ (\ref{gg93}) does fibrewise representation
(\ref{gg82}):
\mar{gg94}\beq
\g_h:M^hX\op\times_X S^h\ar_X S^h. \label{gg94}
\eeq

\begin{remark} \label{gg100} \mar{gg100}
Given a tetrad field, let the Lorentz bundle atlas
$\Psi^h=\{z^h_\iota\}$ (\ref{lat}) of a reduced Lorentz bundle
$L^hX$ gives rise to an atlas $\ol\Psi^h=\{\ol z^h_\iota\}$,
$z^h_\iota =z_h\circ \ol z^h_\iota$, of the principal
$\mathrm{L_s}$-bundle $\wt L^hX$ in the diagram (\ref{gg79}). With
respect to these and associated atlases the representations
(\ref{gg95}) -- (\ref{gg94}) takes a form
\mar{L4'}\beq
\wh h^a=\g_h(h^a)=\g^a,\qquad \wh
dx^\la=\g_h(dx^\la)=h^\la_a(x)\g^a, \label{L4'}
\eeq
where $\g^a$ are Dirac's $\g$-matrices (\ref{41'}) and $h^a$ are
the tetrad coframes (\ref{b3211'}).
\end{remark}

In view of the representations (\ref{gg95}) -- (\ref{gg94}), one
can treat sections of the fibre bundle $S^h$ (\ref{gg78}) as Dirac
spinor fields in the presence of a tetrad field $h$.

However, the representations $\g_h$ and $\g_{h'}$ (\ref{L4'}) for
different tetrad fields $h$ and $h'$ are inequivalent. Indeed,
given elements $t=t_\m dx^\m=t_ah^a=t'_a{h'}^a$ of $T^*X$, their
representations $\g_h$ and $\g_{h'}$ (\ref{L4'}) read
\be
\g_h(t)=t_a\g^a=t_\m h^\m_a\g^a, \qquad \g_{h'}(t)=t'_a\g^a=t_\m
{h'}^\m_a\g^a,
\ee
and lead to non-isomorphic Clifford algebras because
$\g_h(t)\g_h(t')\neq \g_{h'}(t)\g_{h'}(tt')$.

Treating sections of spinor bundles $S^h$ as Dirac spinor fields
in the presence of tetrad fields $h$, one can consider the
composite spinor bundle $S$ (\ref{ggz}) in order to describe the
totality of Dirac spinor fields in the presence of gravitational
field in gauge gravitation theory \cite{book09,sard98,sard11}.  We
agree to call it the \textbf{universal spinor bundle} because,
given a tetrad field $h$, the pull-back $S^h=h^*S\to X$
(\ref{gg78}) of $S_\Si$ (\ref{sp248a}) is a spinor bundle on $X$
which is associated to an $\mathrm{L_s}$-principal bundle $\wt
L^hX$. A universal spinor bundle $S$ is endowed with bundle
coordinates $(x^\la, \si^\m_a,y^A)$, where $(x^\la, \si^\m_a)$ are
bundle coordinates on $\Si_\mathrm{T}$ and $y^A$ are coordinates
on a spinor space $\Psi(4)$. A universal spinor bundle
$S\to\Si_\mathrm{T}$ is a subbundle of the bundle in Clifford
algebras (\ref{gg80}) which is generated by the bundle $M_\Si$
(\ref{gg90}) in Minkowski spaces associated to an L-principal
bundle $LX\to\Si_\mathrm{T}$ (\ref{gg56}). As a consequence, the
fibrewise representation (\ref{gg82}) is defined. It reads
\mar{L7}\beq
\g (dx^\la) =\si^\la_a\g^a. \label{L7}
\eeq

Given the fibrewise Clifford algebra representation (\ref{L7}),
one can introduce a Dirac operator on a spinor bundle $S^h$ for
each tetrad field $h$ as the pull-back of the total Dirac operator
$\cD$ (\ref{gg120}) on the universal spinor bundle $S$ as follows
\cite{book09,sard98,sard11}.

One can show that, due to the splitting (\ref{g13}), any world
connection $\G$ (\ref{B}) on $X$ yields a connection
\mar{b3266}\ben
&& A_\Si = dx^\la\ot(\dr_\la - \frac14
(\eta^{kb}\si^a_\m-\eta^{ka}\si^b_\m)
 \si^\nu_k \G_\la{}^\m{}_\nu I_{ab}{}^A{}_By^B\dr_A) +
 \label{b3266}\\
&& \qquad d\si^\m_k\ot(\dr^k_\m + \frac14 (\eta^{kb}\si^a_\m
-\eta^{ka}\si^b_\m) I_{ab}{}^A{}_By^B\dr_A) \nonumber
\een
on the  spinor bundle $S_\Si\to\Si_\mathrm{T}$ (\ref{sp248a}),
where $I_{ab}$ are the generators (\ref{gg70}). Its pall-back to
$S^h$ is the spin Lorentz connection
\mar{b3212}\beq
\G_s=dx^\la\ot[\dr_\la +\frac14
(\eta^{kb}h^a_\m-\eta^{ka}h^b_\m)(\dr_\la h^\m_k - h^\nu_k
\G_\la{}^\m{}_\nu)I_{ab}{}^A{}_B y^B\dr_A] \label{b3212}
\eeq
associated to the Lorentz connection $\G_h$ (\ref{a3205}) defined
by $\G$ on a reduced Lorentz bundle $L^hX$ . The connection
(\ref{b3266}) yields the vertical covariant differential
\mar{7.10'}\beq
\wt D= dx^\la\ot[y^A_\la- \frac14(\eta^{kb}\si^a_\m
-\eta^{ka}\si^b_\m)(\si^\m_{\la k} -\si^\nu_k
\G_\la{}^\m{}_\nu)L_{ab}{}^A{}_By^B]\dr_A, \label{7.10'}
\eeq
on the universal spinor bundle $S\to X$ (\ref{ggz}). Its
restriction to $S^h\subset S$ recovers the familiar covariant
differential on the spinor bundle $S^h\to X$ relative to the spin
connection (\ref{b3212}). Combining (\ref{L7}) and (\ref{7.10'})
gives the first order differential operator
\mar{gg120}\beq
\cD=\si^\la_a\g^{aB}{}_A[y^A_\la- \frac14(\eta^{kb}\si^a_\m
-\eta^{ka}\si^b_\m)(\si^\m_{\la k} -\si^\nu_k
\G_\la{}^\m{}_\nu)L_{ab}{}^A{}_By^B], \label{gg120}
\eeq
on the universal spinor bundle $S\to X$ (\ref{ggz}). Its
restriction to $S^h\subset S$ is the familiar Dirac operator on a
spinor bundle $S^h$ in the presence of a background tetrad field
$h$ and a general world connection $\G$.

\section{Affine world connections}

The tangent bundle $TX$ of a world manifold $X$ as like as any
vector bundle possesses a natural structure of an affine bundle.
It is associated to a principal bundle $AX$ of oriented affine
frames in $TX$ whose structure group is a general affine group
$GA(4,\mathbb R)$. This structure group is always reducible to a
linear subgroup $GL_4$ since the quotient $GA(4,\mathbb R)/GL_4$
is a vector space $\mathbb R^4$. Treating as an affine bundle, the
tangent bundle $TX$ admits affine connections
\mar{mos033}\beq
A= dx^\la\ot(\dr_\la + \G_\la{}^\al{}_\m(x) \dot x^\m\dot\dr_\al
+\si^\al_\la(x)\dot\dr_\al), \label{mos033}
\eeq
called the \textbf{affine world connections}. They are associated
to principal connections on an affine frame bundle $AX$. Every
affine connection $\G$ (\ref{mos033}) on $TX$ yields a unique
linear connection
\mar{mos032}\beq
\G= dx^\la\ot(\dr_\la + \G_\la{}^\al{}_\m(x) \dot x^\m\dot\dr_\al)
\label{mos032}
\eeq
on $TX$. It is associated to a principal connection on a frame
bundle $LX\subset AX$. Conversely, being equivariant, any
principal connection on a frame bundle $LX\subset AX$ gives rise
to a principal connection on an affine frame bundle $AX$, i.e.,
every linear connection on $TX$ can be seen as the affine one. It
follows that any affine connection $A$ (\ref{mos033}) on the
tangent bundle $TX$ is represented by a sum of the associated
linear connection $\ol\G$ (\ref{mos032}) and a \textbf{soldering
form} $\si=\si^\al_\la(x) dx^\la\ot\dot\dr_\al$ on $TX$, which is
a $(1,1)$-tensor field
\mar{mos035'}\beq
\si=\si^\al_\la(x) dx^\la\ot\dr_\al \label{mos035'} \
\eeq
on $X$ due to the canonical splitting $VTX=TX\times TX$.

In particular, let us consider  the canonical soldering form
$\thh_J$ (\ref{gg11}) on $TX$. Given an arbitrary world connection
$\G$ (\ref{B}) on $TX$, the corresponding affine connection on
$TX$ is a \textbf{Cartan connection}
\be
 A=\G +\thh_X, \qquad
A_\la^\m=\G_\la{}^\m{}_\nu \dot x^\nu +\dl^\m_\la.
\ee

There is a problem of a physical meaning of the tensor field $\si$
(\ref{mos035'}).

In the framework of above mentioned Poincar\'e gauge theory, it is
treated as a non-holonomic frame field or a tetrad field (Remark
\ref{gg7}). This treatment of $\si$ is wrong because a soldering
form and a frame field are different mathematical objects. A frame
field is a (local) section of a principal frame bundle $LX$, while
a soldering form is a global section of the $LX$-associated tensor
bundle
\be
TX\op\ot T^*X=(LX\times \mathrm{Mat}(4, \mathbb R))/GL_4
\ee
whose typical fibre is an algebra $\mathrm{Mat}(4, \mathbb R)$ of
four-dimensional real matrices. It contains a group $GL_4$ which
acts on $\mathrm{Mat}(4, \mathbb R)$ by the adjoint
representation, but not left multiplications.

At the same time, a translation part of an affine connection on
$\mathbb R^3$ characterizes an elastic distortion in gauge theory
of dislocations in continuous media \cite{kad,maly}. By analogy
with this gauge theory, a gauge model of hypothetic deformations
of a world manifold has been developed. They are described by the
translation part $\si$ (\ref{mos035'}) of affine world connections
on $X$ and, in particular, they are responsible for the so called
"fifth force" \cite{sard87,sard90,sardz}.

\end{document}